\documentclass[12pt]{article}

\usepackage{geometry}
\usepackage{amsthm}
\usepackage{amsmath}
  {
	\theoremstyle{plain}
	\newtheorem{assumption}{Assumption}
}
\usepackage{amsfonts}
\usepackage[onehalfspacing]{setspace}
\usepackage[font=small, format=hang]{caption}
\usepackage[authoryear,comma]{natbib}
\usepackage{graphicx}
\usepackage{natbib}
\usepackage{lscape}
\usepackage[latin1]{inputenc}
\usepackage[colorlinks=true,allcolors=blue]{hyperref}
\usepackage{booktabs}
\usepackage{longtable}
\usepackage{ctable}
\usepackage{multirow}
\usepackage{bigstrut}
\usepackage{threeparttable}
\usepackage{epstopdf}
\usepackage{setspace}
\usepackage{float}
\usepackage{amsfonts}
\usepackage{amssymb}
\usepackage{amsthm}
\usepackage{bbm}
\usepackage{pdflscape}

\graphicspath{{../Results/}}
\title{
\begin{LARGE}
Teacher-to-classroom assignment \\ 
and student achievement \end{LARGE} 
}
\begin{small}
\author{Bryan S. Graham\footnote{\normalsize Department of Economics, University of California - Berkeley, and National Bureau of Economic Research, e-mail: \texttt{bgraham@econ.berkeley.edu}}, Geert Ridder\footnote{\normalsize Department of Economics, University of Southern California,  e-mail: \texttt{ridder@usc.edu}}, Petra Thiemann\footnote{\normalsize Department of Economics, Lund University,  and IZA, e-mail: \texttt{petra.thiemann@nek.lu.se}}, Gema Zamarro\footnote{\normalsize Department of Education Reform, University of Arkansas, and CESR, e-mail: \texttt{gzamarro@uark.edu}} \thanks{We thank seminar audiences at the 2017 All-California Econometrics Conference at Stanford University, UC Riverside, NESG, University of Duisburg-Essen, Tinbergen Institute Amsterdam, Lund University, IFN Stockholm, the University of Bristol, and the University of Southern California for helpful feedback. We thank Tommy Andersson,
Kirabo Jackson, Magne Mogstad, Hessel Oosterbeek, Daniele Paserman, and Hashem Pesaran for useful comments and discussions. All the usual disclaimers apply. Financial support for Graham was provided by the National Science Foundation (SES \#1357499, SES \#1851647). Thiemann was affiliated with USC Dornsife INET while working on this paper.}}

\end{small}

\begin{document}

\maketitle

\thispagestyle{empty}

\begin{abstract}
\noindent \singlespacing 
We study the effects of counterfactual teacher-to-classroom assignments on average student achievement in elementary and middle schools in the US. We use the Measures of Effective Teaching (MET) experiment to semiparametrically identify the average reallocation effects (AREs) of such assignments. Our findings suggest that changes in within-district teacher assignments could have appreciable effects on student achievement. Unlike policies which require hiring additional teachers (e.g., class-size reduction measures), or those aimed at changing the stock of teachers (e.g., VAM-guided teacher tenure policies), alternative teacher-to-classroom assignments are resource neutral; they raise student achievement through a more efficient deployment of \emph{existing} teachers. 
\end{abstract}

\noindent {\bf JEL codes:} I20, I21, I24 
\singlespacing
\noindent {\bf Keywords:} teacher quality, teacher assignment, education production, student achievement, average reallocation effects, K--12

\newpage
\setcounter{page}{1}
\onehalfspacing
\section{Introduction}

Approximately four million teachers work in the public elementary
and secondary education system in the United States. These teachers
provide instruction to almost fifty million students, enrolled in
nearly one hundred thousand schools, across more than thirteen thousand
school districts \citep{ConditionOfEducation_NCES2017,DigestOfEducationStatistics_NCES2017}.
Differences in measured student achievement are substantial across
US schools as well as across classrooms within these schools.
Beginning with \citet{Hanushek_AER1971}, a large economics of education
literature attributes cross-classroom variation in student achievement
to corresponding variation in (largely) latent teacher attributes.
These latent attributes are sometimes referred to as teacher quality
or teacher value-added.

The implications of value-added measures (VAM) for education policy
are controversial both within the academy and outside it.\footnote{See, for example, the March 2015 special issue of \emph{Educational
Researcher} on value-added research and policy or the American Statistical
Association's statement on value-added measures \citep{Morganstein_Wasserstein_SPP2014}. } The most contentious applications of VAM involve their use in teacher
tenure and termination decisions \citep[cf.,][]{Chetty_et_al_EN2012,Darling-Hammond_ER2015}.
The premise of such applications is that changes in the stock of existing
teachers \textendash{} specifically rooting out teachers with low
VAMs and retaining those with high ones \textendash{} could lead to
large increases in student achievement and other life outcomes.

In this paper we pose an entirely different question: is it possible
to raise student achievement, without changes to the existing pool
of teachers, by changing who teaches whom? Schools and school districts
are the loci of teacher employment. To keep our analysis policy-relevant,
we therefore focus on the achievement effects of different within-school
and within-district teacher-to-classroom assignment policies. 

For teacher assignment policies to matter, teachers must vary in their
effectiveness in teaching different types of students. For example,
some teachers may be especially good at teaching English language
learners, minority students, or accelerated learners \citep[c.f.,][]{Dee_RESTAT2004, Loeb_et_al_EEPA2014}. Formally educational production must be non-separable in some teacher
and student attributes \citep{Graham_Imbens_Ridder_USC2007,Graham_Imbens_Ridder_QE14,Graham_Imbens_Ridder_JBES18}.

Using data collected in conjunction with the Measures of Effective
Teaching (MET) project, we present experimental evidence of such
non-separabilities. We expand upon standard models of educational production, which typically assume that the effects of teacher and student inputs are separable. Our identification strategy relies on the random assignment of teachers to classrooms in the MET project. Specifically, we study assignments based upon (i) an observation-based, pre-experiment measure of  teaching practice \textendash{} Danielson's \citeyearpar{Danielson2011} Framework for Teaching (FFT) instrument  \textendash{} and (ii) students' and classroom peers' baseline test scores, also measured pre-experiment. 

Alternative teacher assignments change the joint distribution of these variables. When the educational production function is non-separable in these variables, these changes can affect average achievement. Focusing on assignments based on teacher FFT and student baseline achievement, we find that within-district changes in teacher-to-classroom assignments could increase average classroom student achievement by as much as 0.04 standard deviations. This effect corresponds to the estimated difference in average achievement across the teacher-to-classroom assignment which maximizes aggregate achievement versus the one which minimizes it. Perhaps more realistically, comparing the status quo assignment in MET schools \textendash{} which was generated by random assignment of teachers to classrooms \textendash{} with the optimal one generates an estimated increase in average test scores of 0.02 standard deviations.

To benchmark these effects consider a policy which removes the bottom $\tau \times 100$ percent of teachers from classrooms \textendash{} sorted according to their VAM \textendash{} and replaces them with average teachers (i.e., teachers with VAMs of zero). Assuming a Gaussian distribution for teacher value-added, the effect of such an intervention would be to increase the mean of the student test score distribution by
\begin{equation*}
    \left(1-\tau\right)\sigma\frac{\phi\left(\frac{q_{\tau}}{\sigma}\right)}{1-\Phi\left(\frac{q_{\tau}}{\sigma}\right)}
\end{equation*}
standard deviations.\footnote{As is common practice, we abstract from the effects of possible behavioral responses to VAM-guided teacher retention policies. For example, teachers might become demoralized by VAM retention systems and teach less effectively as a result; or they could be motivated to teach more effectively.} Here $\sigma$ corresponds to the standard deviation of teacher value-added and $q_{\tau}$ to its $\tau^{th}$ quantile. 

\citet[Table 2]{Rockoff_AER04} and \citet[Table 6]{Rothstein_QJE2010}
estimate a standard deviation of teacher value-added of between 0.10
and 0.15. Taking the larger estimate and setting $\tau=0.05$  $(0.10)$ generates an expected increase in student test scores of $0.015$ $(0.026)$ standard deviations; smaller than, albeit comparable to, our reallocation effects. We note that replacing five (ten) percent of teachers would be difficult to do in practice. It would be even more difficult to correctly identify the bottom five (ten) percent of teachers (according to VA) and replace them with average ones. We conclude that the achievement effects of teacher assignment policies are meaningful.

In contrast to policies which replace teachers, and/or require the hiring of new ones, the within-school and within-district teacher reassignment
policies we explore in this paper are resource neutral. Districts
require no new teachers or financial resources to implement them. Of course, reassigning teachers across classrooms, and especially across schools within a district, may involve other types of costs. For example, many school districts operate under collective bargaining agreements which give senior teachers partial control over their school assignment \citep[e.g.,][]{Cohen-Vogel_et_al_EEPA2013}.\footnote{ 
Moreover, many teachers have preferences over school and student characteristics. Thus, one might need to pay bonuses to move teachers to certain schools or classrooms (e.g., classrooms with high fractions of disadvantaged students).}

Our decision to use teacher FFT,  students' and peers' baseline test scores as assignment variables was driven by their availability in the MET dataset. Of course, observational measures of teaching practices, such as the FFT, are routinely used for teacher evaluation, promotion, and retention decisions. Its seems reasonable, therefore, to explore whether they might also aid in determining a teacher's classroom assignment. 

It seems likely, however, that assignments based upon a \emph{different} pair of teacher and classroom attributes would lead to greater achievement gains.\footnote{That is, we did not consider the universe of possible assignment variables.} Working with appropriately chosen linear combinations of multiple student and teacher attributes could generate even larger achievement gains. We leave explorations along these lines to future research. 

Readers may reasonably raise questions about the external validity of the results reported below. We believe such skepticism is warranted and encourage readers to view what is reported below as provisional. We do, however, hope our results are sufficiently compelling to motivate additional research and experimentation with teacher-to-classroom assignment policies.

Work by Susanna Loeb and co-authors suggests that US school districts have \emph{de facto} teacher-to-classroom assignment policies \citep{Loeb2011, Grissom2015}. For example, they find that less experienced, minority, and female teachers are more likely to be assigned to predominantly minority classrooms. They also present evidence that principals use teacher assignments as mechanisms for retaining teachers \textendash{} as well as for encouraging less effective teachers to leave \textendash{} and that more experienced teachers exert more influence on classroom assignment decisions. 

Our work helps researchers and policy-makers understand the achievement effects of such policies and the potential benefits of alternative ones. The findings presented below suggest that teacher-to-classroom assignment policies are consequential and that changes to them could meaningfully increase average student achievement.

In addition to our substantive results, we present new identification results for average reallocation effects (AREs). Identification and estimation of AREs under (conditional) exogeneity is considered by \cite{Graham_Imbens_Ridder_QE14,Graham_Imbens_Ridder_JBES18}. Unfortunately these results do not apply directly here. Although teachers were randomly assigned to classrooms as part of the MET experiment, compliance was imperfect. Furthermore some students moved across classrooms after the random assignment of teachers, which raises concerns about bias due to endogenous student sorting. We develop a semiparametric instrumental variables estimator \citep[e.g.,][]{AiChen2003} which corrects for student and teacher non-compliance. Our analysis highlights how complex identification can be in the context of multi-population matching models where agents sort endogenously. 

In independent work, \cite{Aucejo2019} also use the MET data to explore complementarity between student and teacher attributes in educational production. They do not focus on adjusting for teacher and student non-compliance as we do, emphasizing instead an intent-to-treat type analysis. Their paper includes an interesting analysis of the different sub-components of the FFT measure; something not done here. Despite these methodological differences, their results are qualitatively concordant with ours. In earlier work, \cite{Dee_RESTAT2004, Dee2005} uses Project STAR data to study complementarity between teacher and student minority status.

\section{Model and identification}\label{sec:model}

Our goal is to identify the average achievement effects of alternative assignments of teachers to MET classrooms. These are average reallocation effects (AREs), as introduced by \cite{Graham_Imbens_Ridder_USC2007, Graham_Imbens_Ridder_QE14}. The identification challenge is to use the observed MET teacher-to-classroom assignments and outcomes to recover these AREs.

Our analysis is based upon experimentally generated combinations of student and teacher attributes, that is, it exploits the random assignment of teachers to classrooms in the MET experiment.\footnote{To be precise, the randomization was carried out within schools. See Section~\ref{sec:data} for details.} Like in many other field experiments, various deviations from MET's intended protocol complicate our analysis. In this section we outline a semiparametric model of educational production and consider its identification based upon the MET project \emph{as implemented}, using the MET data \emph{as collected}.

It is useful, however, to first explore nonparametric identification of reallocation effects under an ideal implementation of the MET project (henceforth MET \emph{as designed}). Such an approach clarifies how the extra restrictions introduced below allow for the identification of reallocation effects despite non-compliance, attrition, and other deviations from the intended experimental protocol.

\subsection*{Nonparametric identification under ideal circumstances}

Our setting features two populations, one of students and the other of teachers. Each student is distinguished by an observed attribute $X_{i}$, in our case a measure of baseline academic achievement, and an unobserved attribute, say ``student ability,'' $V_{i}$. Similarly, each teacher is characterized by an observed attribute $W_{i}$, in our case an observation-based measure of teaching pedagogy, and an unobserved attribute, say ``teacher quality," $U_{i}$.\footnote{We use ``ability" as a shorthand for latent student attributes associated with higher test scores; likewise we use ``quality" as a shorthand for latent teacher attributes associated with higher test scores.}

Let $i=1,\ldots,N$ index students. Let $C$ be the total number of MET classrooms or equivalently teachers. We define $G_{i}$ to be a $C\times1$ vector
of classroom assignment indicators. The $c^{th}$ element of $G_{i}$ equals one if student $i$ is in classroom $c \in \{1,\dots,C\}$ and zero otherwise.
The indices of student $i's$ peers or classmates are therefore given by the index set \[
p\left(i\right)=\left\{ j\thinspace:\thinspace G_{i}=G_{j},\thinspace i\neq j\right\} .
\]
Next we define the peer average attribute as $\bar{X}_{p\left(i\right)}=\frac{1}{\left|p\left(i\right)\right|}\sum_{j\in p\left(i\right)}X_{j}$
(i.e., the average of the characteristic $X$ across student $i's$ peers). We define $\bar{V}_{p\left(i\right)}$ similarly.

The MET project protocol did not impose any requirements on how students, in a given school-by-grade cell, were divided into classrooms. Evidently schools followed their existing procedures for dividing students within a grade into separate classrooms. An implication of this observation is that the MET experiment implies no restrictions on the joint density
\begin{equation}\label{eq: density_students_peers}
f_{X_{i},V_{i},\bar{X}_{p\left(i\right)},\bar{V}_{p\left(i\right)}}(x,v,\bar{x},\bar{v}),
\end{equation}
beyond the requirement that the density be feasible.\footnote{We cannot, for example, have an assignment which implies all students have above-average peers.} For example, if most schools tracked students by prior test scores, then we would expect $X_{i}$ and $\bar{X}_{p\left(i\right)}$ to positively covary. If, instead, students were randomly assigned to classrooms and hence peers, we would have, ignoring finite population issues, the factorization
\[
f_{X_{i},V_{i},\bar{X}_{p\left(i\right)},\bar{V}_{p\left(i\right)}}(x,v,\bar{x},\bar{v}) = f_{X_{i},V_{i}}(x,v)f_{\bar{X}_{p\left(i\right)},\bar{V}_{p\left(i\right)}}(\bar{x},\bar{v})
.\footnote{Indeed additional restrictions on the second density to the right of the equality above would also hold; see \citealp{Graham_Imbens_Ridder_NBER10}, and \citealp{Graham_HSE11}, for additional discussion and details.}\] Our analysis allows for arbitrary dependence between own and peer attributes, both observed and unobserved, and consequently is agnostic regarding the protocol used to group students into classrooms.

Two implications of this agnosticism are (i) our analysis is necessarily silent about the presence and nature of any peer group effects, and (ii) it is likely that more complicated policies, involving \emph{simultaneously} regrouping students into new classes \emph{and} reassigning teachers to them, could raise achievement by more than what is feasible via reassignments of teachers to \emph{existing} classrooms alone, which is the class of policies we consider.\footnote{Learning about the effects of policies which simultaneously regroup students and reassign teachers would require double randomization. See \cite{Graham_EM08} for an empirical example and \cite{Graham_Imbens_Ridder_NBER10,Graham_Imbens_Ridder_JBES18} for a formal development.}

Although nothing about the MET protocol generates restrictions on the joint density \eqref{eq: density_students_peers}, random assignment of teachers to classrooms \textendash{} however formed \textendash{} ensures that
\begin{equation}\label{eq: density_students_peers_teachers}
f_{X_{i},V_{i},\bar{X}_{p\left(i\right)},\bar{V}_{p\left(i\right)},W_{i},U_{i}}(x,v,\bar{x},\bar{v},w,u)=f_{X_{i},V_{i},\bar{X}_{p\left(i\right)},\bar{V}_{p\left(i\right)}}(x,v,\bar{x},\bar{v})f_{W_{i},U_{i}}(w,u).
\end{equation}
Here $W_{i}$ and $U_{i}$ denote the observed and unobserved attributes of the teacher assigned to the classroom of student $i$. A perfect implementation of MET as designed would ensure that student and teacher attributes vary independently of each other. Our research design is fundamentally based upon restriction \eqref{eq: density_students_peers_teachers}.

Let $Y_{i}$ be an end-of-year measure of student achievement, generated according to
\begin{equation}\label{eq: nonparametric_production}
    Y_{i}=g\left(X_{i},\bar{X}_{p\left(i\right)},W_{i},V_{i},\bar{V}_{p\left(i\right)},U_{i}\right).
\end{equation}
Other than the restriction that observed and unobserved peer attributes enter as means, equation~\eqref{eq: nonparametric_production} imposes no restrictions on educational production.\footnote{A fully nonparametric model would allow achievement to vary with any exchangeable function of peer attributes; see \cite{Graham_Imbens_Ridder_NBER10} for more details.} Under restriction \eqref{eq: density_students_peers_teachers} the conditional mean of the outcome given observed own, peer, and teacher attributes equals
\begin{equation}\label{eq: average_match_function}
\begin{split}
\mathbb{E}\left[\left.Y_{i}\right|X_{i}=x,
           \bar{X}_{p\left(i\right)} = \bar{x},W_{i}=w\right] & = \iiint \left[ g\left(x,\bar{x},w,v,\bar{v},u\right)f_{\left.V_{i},\bar{V}_{p\left(i\right)}\right|X_{i},X_{p\left(i\right)}}\left(\left.v,\bar{v}\right|x,\bar{x}\right) \right. \\ 
           & \left. \times f_{\left.U_{i}\right|W_{i}}\left(\left.u_{i}\right|w_{i}\right) \right] \mathrm{d}v\mathrm{d}\bar{v}\mathrm{d}u \\
           & = m^{\mathrm{amf}}\left(x,\bar{x},w\right).
\end{split}
\end{equation}
Equation \eqref{eq: average_match_function} coincides with (a variant of) the Average Match Function (AMF) estimand discussed by \cite{Graham_Imbens_Ridder_QE14,Graham_Imbens_Ridder_JBES18}. The AMF can be used to identify AREs. Our setting \textendash{} which involves multiple students being matched to a single teacher \textendash{} is somewhat more complicated than the one-to-one matching settings considered by \cite{Graham_Imbens_Ridder_QE14,Graham_Imbens_Ridder_JBES18}. One solution to this ``problem" would be to average equation \eqref{eq: nonparametric_production} across all students in the same classroom and work directly with those averages. As will become apparent below, however, working with a student-level model makes it easier to deal with non-compliance and attrition, which have distinctly student-level features. It also connects our results more directly with existing empirical work in the economics of K-to-12 education, where student-level modelling predominates, and results in greater statistical power.

The decision to model outcomes at the student level makes the analysis of teacher re-assignments a bit more complicated, at least superficially. To clarify the issues involved it is helpful to consider an extended example. Assume there are two \emph{types} of students, $X_i \in \{0,1\}$, and two \emph{types} of teachers, $W_i \in \{0,1\}$. For simplicity assume that the population fractions of type $X_i=1$ students and type $W_i=1$ teachers both equal one-half, that is, half of the students are of type 1 and half of the students are taught by a teacher of type 1. Assume, again to keep things simple, that  classrooms consist of three students each. 

Table \ref{table: toy_example} summarizes this basic set-up. Column 1 lists classroom types. For example, a $000$ classroom consists of all type-0 students.  There are four possible classroom types, each assumed to occur with a frequency of one-fourth.\footnote{Within classrooms students are exchangeable.} The status quo mechanism for grouping students into classrooms induces a joint distribution of own and peer average attributes. This joint distribution is given in the right-most column of Table \ref{table: toy_example}. For instance, $\frac{1}{4}$ (3 out of 12) of the students are in a classroom with two type-0 peers, so that $f_{X_{i},\bar{X}_{p\left(i\right)}}(0,0)=\frac{1}{4}$, and $\frac{1}{6}$ (2 out of 12) of the students are in a classroom with one type-0 and one type-1 peer, so that $f_{X_{i},\bar{X}_{p\left(i\right)}}(0,\frac{1}{2})=\frac{1}{6}$.  The MET experiment implies no restrictions on the joint density $f_{X_{i},\bar{X}_{p\left(i\right)}}(x,\bar{x})$, consequently we only consider policies which leave it unchanged.

Next assume, as was the case in the MET experiment, that under the status quo teachers are randomly assigned to classrooms. This induces the conditional distribution of $W_i$ given $X_i$ and $\bar{X}_{p\left(i\right)}$ reported in column 2 of Table~\ref{table: toy_example}. Of course, from this conditional distribution, and the marginal for $X_i$ and $\bar{X}_{p\left(i\right)}$, we can recover the joint distribution of own type, peer average type, and teacher type (i.e., of $X_i$, $\bar{X}_{p\left(i\right)}$ and $W_{i}$).

\begin{table}[t!]
\caption{Feasible teacher reassignments}
\label{table: toy_example}
\centering
 \begin{tabular}{||c c c c c c||} 
 \hline
& Status quo & Counterfactual & & & \\
 Classroom Type & $\Pr\left(\left.W_{i}=1\right|X_i,\bar{X}_{p\left(i\right)}\right)$ 
 & $\Pr\left(\left.\tilde{W}_{i}=1\right|X_i,\bar{X}_{p\left(i\right)}\right)$ & $X_{i}$ & $\bar{X}_{p\left(i\right)}$ 
 & $f\left(X_i,\bar{X}_{p\left(i\right)}\right)$\\ 
 
 \hline\hline
 
 000 & \( \frac{1}{2} \) & \( \frac{1}{3} \) & 0 & 0 
     & \( \frac{1}{4} \)  \\ 
 
     &  &  & 0 & 0 & \( \frac{1}{4} \) \\
 
     &  &  & 0 & 0 & \( \frac{1}{4} \)  \\
 
 \hline
 
 001 & \( \frac{1}{2} \) & \( \frac{1}{2} \) & 0 
     & \( \frac{1}{2} \) & \( \frac{1}{6} \) \\
     
     &  &  & 0 & \( \frac{1}{2} \) & \( \frac{1}{6} \) \\
    
     &  &  & 1 & 0 & \( \frac{1}{12} \) \\
 \hline
 
 011 & \( \frac{1}{2} \) & \( \frac{1}{2} \) & 0 & 1 
     & \( \frac{1}{12} \) \\ 
 
     &  &  & 1 & \( \frac{1}{2} \) & \( \frac{1}{6} \) \\
  
     &  &  & 1 & \( \frac{1}{2} \) & \( \frac{1}{6} \) \\ 
  
 \hline
 
 111 & \( \frac{1}{2} \) & \( \frac{2}{3} \) & 1 & 1 
     & \( \frac{1}{4} \)\\   
 
     &  &  & 1 & 1 & \( \frac{1}{4} \) \\ 
     
     &  &  & 1 & 1 & \( \frac{1}{4} \) \\
     
\hline
     
\end{tabular}

\caption*{ \footnotesize \textit{Note:} The population fraction of type $X_i=1$ students is \( \frac{1}{2} \) and that of type $W_i=1$ teachers is also \( \frac{1}{2} \). Classrooms of three students each are formed, such that the frequency of each of the four possible classroom configurations is \( \frac{1}{4} \) in the population of classrooms of size 3. Under the status quo teachers are assigned to classrooms at random; in the counterfactual teachers are assigned more assortatively. See the main text for more information.}

\end{table}

Now consider the AMF: $m^{\mathrm{amf}}\left(x,\bar{x},w\right)$. Consider the subpopulation of students with $X_{i}=1$ and $\bar{X}_{p\left(i\right)}=\frac{1}{2}$. Inspecting Table \ref{table: toy_example}, this subpopulation represents \( \frac{1}{6} \) of all students (right-most column of Table \ref{table: toy_example}). If we assign to students in this subpopulation a teacher of type $W_{i}=1$, then the expected outcome coincides with $m^{\mathrm{amf}}\left(1,\frac{1}{2},1\right)$. Under random assignment of teachers the probability of assigning a type-1 teacher is the same for all subpopulations of students. 

Finally consider a counterfactual assignment of teachers to classrooms. Since we leave the composition of classrooms unchanged, $f_{X_{i},\bar{X}_{p\left(i\right)}}(x,\bar{x})$ is left unmodified. The counterfactual assignment therefore corresponds to a conditional distribution for teacher type, $\tilde{f}_{\left. \tilde{W}_{i} \right| X_{i},\bar{X}_{p\left(i\right)}} \left(\left. w \right| x,\bar{x}\right)$ which satisfies the feasibility condition:
\begin{equation}\label{eq: feasibility_constraint}
\begin{split}
\iint\tilde{f}_{\left.\tilde{W}_{i}\right|X_{i},\bar{X}_{p\left(i\right)}}\left(\left.w\right|x,\bar{x}\right)f_{X_{i},\bar{X}_{p\left(i\right)}}\left(x,\bar{x}\right)\mathrm{d}x\mathrm{d}\bar{x}=f\left(w\right)
\end{split}
\end{equation}
for all $w \in \mathbb{W}$. Here  $\tilde{f}$ denotes a counterfactual distribution, while  $f$ denotes a status quo one. We use $\tilde{W}_{i}$ to denote an assignment  from the counterfactual distribution. Note that by feasibility of an assignment $\tilde{W}_{i}\stackrel{D}{=}W_{i}$ marginally, but will differ conditional on student attributes. Condition \eqref{eq: feasibility_constraint}, as discussed by \cite{Graham_Imbens_Ridder_QE14}, allows for degenerate conditional distributions, as might occur under a perfectly positive assortative matching.

Average achievement under a counterfactual teacher-to-classroom assignment equals:
\begin{equation}\label{eq: ARE_identified}
    \beta^{\mathrm{are}}\left(\tilde{f}\right)=\iint\left[\int m^{\mathrm{amf}}\left(x,\bar{x},w\right)\tilde{f}_{\left.\tilde{W}_{i}\right|X_{i},\bar{X}_{p\left(i\right)}}\left(\left.w\right|x,\bar{x}\right)\mathrm{d}w\right]f_{X_{i},\bar{X}_{p\left(i\right)}}\left(x,\bar{x}\right)\mathrm{d}x\mathrm{d}\bar{x}.
\end{equation}
Since all the terms to the right of the equality are identified, so too is the ARE. Conceptually we first \textendash{} see the inner integral in equation \eqref{eq: ARE_identified} \textendash{} compute the expected outcome in each type of classroom (e.g., $X_{i}=x$ and $\bar{X}_{p\left(i\right)}=\bar{x}$) given its new teacher assignment (e.g., to type $\tilde{W}_{i}=w$). We then \textendash{} see the outer two integrals in equation \eqref{eq: ARE_identified} \textendash{} average over the status quo distribution of $X_{i},\bar{X}_{p\left(i\right)}$, which is left unchanged. This yields average student achievement under the new assignment of teachers to classrooms.

In addition to the feasibility condition \eqref{eq: feasibility_constraint} we need to also rule out allocations that assign different teachers to students in the same classrooms. Note that $m^{\mathrm{amf}}\left(x,\bar{x},w\right)$ is the average outcome for the subpopulation of students of type $X_i=x$ with peers $\bar{X}_{p(i)}=\bar{x}$. For example, in Table \ref{table: toy_example} classroom 001 has students from two subpopulations so defined. Assignment of teachers to subpopulations of students opens up the possibility that a classroom is assigned to teachers of different types for its constituent subgroups of students. If, as indicated in Table \ref{table: toy_example}, the teacher-type assignment probability is the same for all subpopulations of students represented in a classroom, then the ARE in equation \eqref{eq: ARE_identified} coincides with one based on direct assignment of teachers to classrooms. This implicit restriction on teacher assignments provides a link between models for individual outcomes and classroom-level reallocations.

\subsection*{Semiparametric identification under MET as implemented}

In the MET experiment as implemented not all teachers and students appear in their assigned classrooms. This occurs both due to attrition (e.g., when a student changes schools prior to follow-up) as well as actual non-compliance (e.g., when a teacher teaches in a classroom different from their randomly assigned one).

In this section we describe our approach to identifying AREs in MET as implemented. Relative to the idealized analysis of the previous subsection we impose two types of additional restrictions. First, we work with a semiparametric, as opposed to a nonparametric, educational production function. Second, we make behavorial assumptions regarding the nature of non-compliance. Both sets of assumptions are (partially) testable.

\subsubsection*{Educational production function}
Our first set of restrictions involve the form of the educational production function. A key restriction we impose is that \emph{unobserved} student, peer, and teacher attributes enter separably. Although this assumption features in the majority of economics of education empirical work \citep[e.g.,][]{Chetty_et_al_AER2014a,Chetty_et_al_AER2014b}, it is restrictive. We also discretize the observed student and teacher attribute. This allows us to work with a parsimonously parameterized educational production function that nevertheless accommodates complex patterns of complementarity between student and teacher attributes. Discretization also allows us to apply linear programming methods to study counterfactual assigments \citep[c.f.,][]{Graham_Imbens_Ridder_USC2007, Bhattacharya_JASA2009}.

Specifically we let $X_{i}$ be a vector of indicators for each of $K$ ``types" of students. Types correspond to intervals of baseline test scores. Our preferred specification works with $K=3$ types of students: those with low, medium, and high baseline test scores.\footnote{Precise variable definitions are given below.} In this case $X_{i}$ is a $2 \times 1$ vector of dummies for whether student $i$'s baseline test score was in the medium or high range (with the low range being the omitted group). This definition of $X_{i}$ means that $\bar{X}_{p(i)}$ equals the $2 \times 1$ vector of fractions of peers in the medium and high baseline categories (with the fraction low range omitted).

We discretize the distribution of the teacher attribute similarly: $W_{i}$ is a vector of indicators for $L$ different ranges of FFT scores. In our preferred specification we also work with $L=3$ types of teachers: those with low, medium, and high FFT scores. Hence $W_{i}$ is again a $2 \times 1$ vector of dummies for whether the teacher of student $i$'s FFT score was in the medium or high range (with the low range again being the omitted group). 

We assess the sensitivity of our results to coarser and finer discretizations of the baseline test score and FFT distributions. Specifically we look at $K=L=2$ and $K=L=4$ dicretizations.

We posit that end-of-school year achievement for student $i$ is generated according to
\begin{equation}\label{eq: outcome_model}
\begin{split}
Y_{i} &= \underset{\text{Student Ability}}{\underbrace{\alpha+X_{i}'\beta+V_{i}}}
+\underset{\text{Peer Effect}}{\underbrace{\bar{X}_{p\left(i\right)}'\gamma+\rho\bar{V}_{p\left(i\right)}}}+\underset{\text{Teacher Quality}}{\underbrace{W_{i}'\delta+U_{i}}}\\
     &+\underset{\text{Student-Peer Complementarity}}{\underbrace{\left(X_{i}\otimes\bar{X}_{p\left(i\right)}\right)'\zeta}} 
+\underset{\text{\text{Student-Teacher Complementarity}}}{\underbrace{\left(X_{i}\otimes W_{i}\right)'\eta+\left(W_{i}\otimes\bar{X}_{p\left(i\right)}\right)'\lambda}}
\end{split}
\end{equation}
Observe that \textendash{} as noted above \textendash{} own, $V_i$, peer, $\bar{V}_{p(i)}$, and teacher, $U_{i}$, unobservables enter equation \eqref{eq: outcome_model} additively. The labelled grouping of terms in equation \eqref{eq: outcome_model} highlights the flexibility of our model relative to those typically employed by researchers. As in traditional models, observed and unobserved student and teacher attributes are posited to directly influence achievement. We add to this standard set-up the possibility of complementarity between own and peer attributes, and complementarity between own and teacher attributes. Additionally our model allows for both observed and \emph{unobserved} peer attributes to influence achievement.

Conditional on working with a discrete student and teacher type space, equation \eqref{eq: outcome_model} is unrestrictive in how own and teacher attributes interact to generate achievement. In contrast, equation \eqref{eq: outcome_model} restricts the effect of peers' observed composition on the outcome. Partition $\zeta=\left(\zeta_{1},\dots,\zeta_{K-1}\right)$ and similarly partition $\lambda=\left(\lambda_{1},\dots,\lambda_{L-1}\right)$. The $(K-1)\times1$ gradient of student $i$'s outcome with respect to peer composition is
\begin{equation}
\frac{\partial Y_{i}}{\partial\bar{X}_{p\left(i\right)}}=\gamma+\sum_{k=1}^{K-1}X_{ki}\zeta_{k}+\sum_{l=1}^{L-1}W_{li}\lambda_{l},
\end{equation}
which is constant in $\bar{X}_{p\left(i\right)}$, although varying heterogenously with student and teacher type. Put differently convexity/concavity in $\bar{X}_{p\left(i\right)}$ is ruled out by equation \eqref{eq: outcome_model}. It should be noted that the MET data, in which the assignment of peers is not random, are not suitable for estimating peer effects (non-linear or otherwise).

For completeness we also include the interaction of teacher type with peer composition \textendash{} the $\left(W_{i}\otimes\bar{X}_{p\left(i\right)}\right)$ regressor in equation \eqref{eq: outcome_model} \textendash{}  although $\lambda$ is poorly identified in practice.\footnote{Admittedly, the $\left(W_{i}\otimes\bar{X}_{p\left(i\right)}\right)$ term is not entirely straightforward to interpret; however it seemed ad hoc to include some second-order interactions while a priori excluding others. We also report specifications which exclude this term below (as its coefficient is always insignificantly different from zero).} Due to our limited sample size, we do not include the third order interactions of own, peer, and teacher types.\footnote{The educational production function in equation \eqref{eq: outcome_model} includes $J=\mathrm{dim}(\alpha)+\mathrm{dim}(\beta)+\mathrm{dim}(\gamma)+\mathrm{dim}(\delta)+\mathrm{dim}(\zeta)+\mathrm{dim}(\eta)+\mathrm{dim}(\lambda)=1+2+2+2+4+4+4=19$ parameters. A fully interacted model would introduce $8$ additional parameters, for $3 \times 3 \times 3 = 27$ in total. Even this model would not be fully flexible since, although constructed from averages of binary indicators, $\bar{X}_{p(i)}$ is not binary-valued. A more flexible model would therefore also include, for example, interactions of $X_i$ with the squares of the elements of $\bar{X}_{p(i)}$ (and so on.)}

Relative to a standard ``linear-in-means" type model typically fitted to datasets like ours \citep[e.g.,][]{HanushekKainRivkin2004}:
\begin{equation}\label{eq: linear-in-means}
    Y_{i}=\alpha+X_{i}'\beta+\bar{X}_{p\left(i\right)}'\gamma+W_{i}'\delta+V_{i}+U_{i},
\end{equation}
equation \eqref{eq: outcome_model} is rather flexible. It allows for rich interactions in observed own, peer, and teacher attributes and is explicit in that both observed \emph{and} unobserved peer attributes may influence own achievement. The ``linear-in-means" model \eqref{eq: linear-in-means} presumes homogenous effects and does not explicitly incorporate unobserved peer attributes.\footnote{\cite{Manski_ReStud93} did allow for unobserved peer attributes as did \cite{Graham_EM08}; but these cases are exceptional.}

As mentioned above, a student's assigned teacher and peers may deviate from her realized ones due to attrition and non-compliance. To coherently discuss our assumptions about these issues we require some additional notation.
Let $W_{i}^{*}$ and $\bar{X}_{p^*\left(i\right)}$ denote student
$i's$ \emph{assigned} teacher and peer attribute (here $p^{*}\left(i\right)$
is the index set of $i's$ \emph{assigned} classmates). Random assignment of teachers to classrooms ensures that a student's \emph{assigned} teacher's attributes are independent of her own unobservables:
\begin{equation}\label{met: MET_implication_1}
\mathbb{E}\left[\left.V_{i}\right|X_{i},\bar{X}_{p^{*}(i)},W_{i}^{*}\right] =\mathbb{E}\left[\left.V_{i}\right|X_{i},\bar{X}_{p^{*}\left(i\right)}\right]\stackrel{def}{\equiv}g_{1}\left(X_{i},\bar{X}_{p^{*}\left(i\right)}\right).
\end{equation}
Here $g_{1}(x,\bar{x})$ is unrestricted. Under double randomization, with students additionally grouped into classes at random, we would have the further restriction
\[
\mathbb{E}\left[\left.V_{i}\right|X_{i},\bar{X}_{p^{*}\left(i\right)}\right]=\mathbb{E}\left[\left.V_{i}\right|X_{i}\right].
\]
However, since the MET experiment placed no restrictions on how students were grouped into classrooms, we cannot rule out the possibility that a student's peer characteristics, $\bar{X}_{p^{*}\left(i\right)}$, predict her own unobserved ability, $V_i$. Consequently our data are necessarily silent about the presence and nature of any peer group effects in learning. This limitation does not limit our ability to study the effects of teacher reallocations, because we leave the student composition of classrooms \textendash{} and hence the ``peer effect" \textendash{} fixed in our counterfactual experiments.

Finally, even with double randomization, we would still have $\mathbb{E}\left[\left.V_{i}\right|X_{i}\right]\neq 0$. Observed and unobserved attributes may naturally covary in any population (e.g., for example, average hours of sleep, which is latent in our setting, plausibly covaries with baseline achievement and also influences the outcome). Such covariance is only a problem if, as is true in the traditional program evaluation setting, the policies of interest induce changes in the marginal distribution of $X_i$ \textendash{} and hence the joint distribution of $X_i$ and $V_i$.\footnote{A prototypical example is a policy which increases years of completed schooling. Such a policy necessarily changes the joint distribution of schooling and unobserved labor market ability relative to its status quo distribution.} This is not the case here: any reallocations leave the joint distribution of $X_{i}$ and $V_{i}$ unchanged.

The MET protocol also ensures that assigned peer unobservables, $\bar{V}_{p^*(i)}$, are independent of the observed attributes of one's assigned teacher:
\begin{equation}\label{met: MET_implication_2}
\mathbb{E}\left[\left.\bar{V}_{p^{*}(i)}\right|X_{i},\bar{X}_{p^{*}\left(i\right)},W_{i}^{*}\right]=\mathbb{E}\left[\left.\bar{V}_{p^{*}(i)}\right|X_{i},\bar{X}_{p^{*}\left(i\right)}\right] 
\stackrel{def}{\equiv}g_{2}\left(X_{i},\bar{X}_{p^{*}\left(i\right)}\right),
\end{equation}
with $g_{2}\left(X_{i},\bar{X}_{p^{*}\left(i\right)}\right)$, like  $g_{1}\left(X_{i},\bar{X}_{p^{*}\left(i\right)}\right)$, unrestricted.

Random assignment of teachers to classrooms also ensures independence of the unobserved attribute of a student's \emph{assigned} teacher and observed student and peer characteristics:
\begin{equation}\label{met: MET_implication_3}
\mathbb{E}\left[\left.U_{i}^{*}\right|X_{i},\bar{X}_{p^{*}\left(i\right)},W_{i}^{*}\right]=\mathbb{E}\left[\left.U_{i}^{*}\right|W_{i}^{*}\right]=0.
\end{equation}
The second equality is just a normalization.\footnote{Reallocations leave the joint distribution of $U_{i}$ and $W_{i}$ unchanged, so we are free to normalize this mean to zero.}

Under MET \emph{as designed} we could identify AREs using equations \eqref{met: MET_implication_1}, \eqref{met: MET_implication_2}, and \eqref{met: MET_implication_3}. To see this let, as would be true under perfect compliance, $W_{i}=W_{i}^{*}$ and $\bar{X}_{p\left(i\right)}=\bar{X}_{p^{*}\left(i\right)}$ for all $i=1,\dots,N$. Using equations \eqref{met: MET_implication_1}, \eqref{met: MET_implication_2}, and \eqref{met: MET_implication_3} yields, after some manipulation, the partially linear regression model \citep[e.g.,][]{Robinson_EM88}:
\begin{equation}\label{eq: plm_perfect_experiment}
    Y_{i}=W_{i}'\delta+\left(X_{i}\otimes W_{i}\right)'\eta+\left(W_{i}\otimes\bar{X}_{p\left(i\right)}\right)'\lambda+h\left(X_{i},\bar{X}_{p\left(i\right)}\right)+A_{i}
\end{equation}
with $\mathbb{E}\left[\left.A_{i}\right|X_{i},\bar{X}_{p\left(i\right)},W_{i}\right]=0$ for
\begin{equation}
    A_{i}\overset{def}{\equiv}\left[V_{i}-g_{1}\left(X_{i},\bar{X}_{p\left(i\right)}\right)\right]+\rho\left[\bar{V}_{p\left(i\right)}-g_{2}\left(X_{i},\bar{X}_{p\left(i\right)}\right)\right]+U_{i},
\end{equation}
and where the nonparametric regression component equals
\begin{equation}
    h\left(X_{i},\bar{X}_{p\left(i\right)}\right)\overset{def}{\equiv}\alpha+X_{i}'\beta+g_{1}\left(X_{i},\bar{X}_{p\left(i\right)}\right)+\bar{X}_{p\left(i\right)}'\gamma+\rho g_{2}\left(X_{i},\bar{X}_{p\left(i\right)}\right)+ \left(X_{i}\otimes\bar{X}_{p\left(i\right)}\right)'\zeta.
\end{equation}
Note, even under this perfect experiment, we cannot identify $\beta$, $\gamma$, and $\zeta$; these terms are confounded by $g_{1}\left(X_{i},\bar{X}_{p\left(i\right)}\right)$ and $g_{2}\left(X_{i},\bar{X}_{p\left(i\right)}\right)$ and hence absorbed into the nonparametric component of the regression model. This lack of identification reflects the inherent inability of the MET experiment to tell us anything about peer group effects. Nevertheless, as detailed below, knowledge of $\delta$, $\eta$, and $\lambda$ is sufficient to identify the class of reallocation effects we focus upon.

\subsubsection*{Patterns of non-compliance}
Unfortunately, we do not observe student outcomes under full compliance. Non-compliance may induce correlation between $A_i$ and $X_i$, $\bar{X}_{p\left(i\right)}$ and $W_i$ in regression model \eqref{eq: plm_perfect_experiment}. Our solution to this problem is to construct instrumental variables for \emph{observed} teacher and peer attributes, $W_i$ and $\bar{X}_{p\left(i\right)}$ \textendash{} which necessarily reflect any non-compliance and attrition on the part of teachers and students \textendash{} from the \emph{assigned} values, $W^{*}_{i}$ and $\bar{X}_{p^{*}\left(i\right)}$.

Rigorously justifying this approach requires imposing restrictions on how, for example, realized and assigned teacher quality relate to one another. The first assumption we make along these lines is:
\begin{assumption}\label{ass:key_ass_1} \textsc{(Idiosyncratic Teacher Deviations)}
	\begin{equation}
	\mathbb{E}\left[\left.U_{i}-U_{i}^{*}\right|X_{i},\bar{X}_{p^{*}\left(i\right)},W_{i}^{*}\right]=0.\label{eq: key_ass_1}
	\end{equation}
\end{assumption}
Assumption \ref{ass:key_ass_1} implies that the \emph{difference} between \emph{realized} and \emph{assigned}
(unobserved) teacher ``quality'' cannot be predicted by own and assigned
peer and teacher observables. While Assumption \ref{ass:key_ass_1} is not directly testable, we can perform the following plausibility test. Let $R_{i}-R_{i}^{*}$ be the difference
between the realized and assigned value of some observed teacher
attribute other than $W_i$ (e.g., years of teaching experience). Under equation~\eqref{eq: key_ass_1}, if we compute the OLS fit of this difference onto $1$,
$X_{i}$, $W_{i}^{*}$, and $\bar{X}_{p^{*}\left(i\right)}$, a test for the joint significance of the non-constant regressors should accept the null of no effect. Finding that, for example, students \emph{assigned} to classrooms with low average peer prior-year achievement, tend to \emph{move into} classrooms with more experienced teachers suggests that Assumption \ref{ass:key_ass_1} may be implausible.

Assumption \ref{ass:key_ass_1} and equation \eqref{met: MET_implication_3} above yield the mean independence restriction
\begin{equation}\label{eq: identification_result_1}
\mathbb{E}\left[\left.U_{i}\right|X_{i},W_{i}^{*},\bar{X}_{p^{*}\left(i\right)}\right]=0.
\end{equation}
This equation imposes restrictions on the unobserved attribute of student $i$'s \emph{realized} teacher. It is this latent variable which drives the student outcome actually observed.

Our second assumption involves the relationship between the unobserved attributes of a student's assigned peers and those of her realized peers. These two variables will differ if some students switch out of their assigned classrooms.
\begin{assumption}\label{ass:key_ass_2} \textsc{(Conditionally Idiosyncratic Peer Deviations)}
	\begin{equation}
	\mathbb{E}\left[\left.\bar{V}_{p(i)}-\bar{V}_{p^{*}(i)}\right|X_{i},\bar{X}_{p^{*}\left(i\right)},W_{i}^{*}\right]=\mathbb{E}\left[\left.\bar{V}_{p(i)}-\bar{V}_{p^{*}(i)}\right|X_{i},\bar{X}_{p^{*}\left(i\right)}\right]\label{eq: key_ass_2}
	\end{equation}
\end{assumption}
Assumption \ref{ass:key_ass_2} implies that the \emph{difference} between \emph{realized } and \emph{assigned} unobserved peer ``quality'' cannot be predicted by assigned teacher
observables. We do allow for these deviations to covary with a student's type and the assigned composition of her peers. Assumption \ref{ass:key_ass_2} and equation \eqref{met: MET_implication_2} yield a second mean independence restriction of
\begin{equation}\label{eq: identification_result_2}
\mathbb{E}\left[\left.\bar{V}_{p(i)}\right|W_{i}^{*},X_{i},\bar{X}_{p^{*}\left(i\right)}\right]=g^{*}_{2}\left(X_{i},\bar{X}_{p^{*}\left(i\right)}\right),
\end{equation}
where $g^{*}_{2}\left(X_{i},\bar{X}_{p^{*}\left(i\right)}\right) \stackrel{def}{\equiv} \mathbb{E}\left[\left.\bar{V}_{p(i)}-\bar{V}_{p^{*}(i)}\right|X_{i},\bar{X}_{p^{*}\left(i\right)}\right] + g_{2}\left(X_{i},\bar{X}_{p^{*}\left(i\right)}\right)$ is unrestricted.
We can also pseudo-test Assumption \ref{ass:key_ass_2} using observed peer attributes. Finding, for example, that \textendash{} conditional on own type, $X_i$, and assigned peers' average type, $\bar{X}^{*}_{p(i)}$ \textendash{} assigned teacher quality, $W^{*}_{i}$, predicts differences between the realized and assigned values of other observed peer attributes provides evidence against Assumption \ref{ass:key_ass_2}.

The experiment-generated restrictions \textendash{} equations \eqref{met: MET_implication_1}, \eqref{met: MET_implication_2}, and \eqref{met: MET_implication_3} above \textendash{} in conjunction with our two (informally testable) assumptions about deviations from the experiment protocol \textendash{} Assumptions \ref{ass:key_ass_1} and \ref{ass:key_ass_2} above \textendash{} together imply the following conditional moment restriction:
\begin{align}
\mathbb{E}\left[\left.U_{i}+V_{i}+\rho\bar{V}_{p\left(i\right)}\right|W_{i}^{*},X_{i},\bar{X}_{p^{*}\left(i\right)}\right] & =g_{1}\left(X_{i},\bar{X}_{p^{*}\left(i\right)}\right)+\rho g^{*}_{2}\left(X_{i},\bar{X}_{p^{*}\left(i\right)}\right).\label{eq: conditional_moment_restriction}
\end{align}
We wish to emphasize two features of restriction \eqref{eq: conditional_moment_restriction}. First, the conditioning variables are \emph{assigned} peer and teacher attributes, not their realized counterparts. This reflects our strategy of using assignment constructs as instruments. Second, any function of $W^{*}_{i}$, as well as interactions of such functions with functions of $X_i$ and $\bar{X}_{p^{*}\left(i\right)}$ do not predict the composite error $U_{i}+V_{i}+\rho\bar{V}_{p\left(i\right)}$ conditional on $X_i$ and $\bar{X}_{p^{*}\left(i\right)}$; hence such terms are valid instrumental variables.

More specifically we redefine $h$ to equal
\begin{equation}
\begin{split}
    h\left(X_{i},\overline{X}_{p^{*}\left(i\right)}, \overline{X}_{p(i)}\right)\overset{def}{\equiv} & \alpha+X_{i}'\beta+g_{1}\left(X_{i},\overline{X}_{p^{*}\left(i\right)}\right)  + \overline{X}_{p(i)}' \gamma\\ 
     & + \rho g_{2}^*\left(X_{i},\overline{X}_{p^{*}\left(i\right)}\right)+ \left(X_{i}\otimes\overline{X}_{p(i)}\right )'\zeta
\label{eq: h()}  
\end{split}    
\end{equation}

\noindent and $A_{i}$ to equal
\begin{equation}
A_{i} \stackrel{def}{\equiv}  \left(V_{i}-g_{1}\left(X_{i},\overline{X}_{p^{*}\left(i\right)}\right)\right)+\rho\left(\overline {V}_{p\left(i\right)}-g^{*}_{2}\left(X_{i},\overline{X}_{p^{*}\left(i\right)}\right)\right) + U_{i}.
\label{eq: A}
\end{equation}
Equations \eqref{eq: outcome_model}, \eqref{eq: h()}, and \eqref{eq: A} yield an outcome equation of
\begin{equation}\label{eq: plm_endogenous_version}
    Y_{i}=W_{i}'\delta+\left(X_{i}\otimes W_{i}\right)'\eta+\left(W_{i}\otimes\bar{X}_{p\left(i\right)}\right)'\lambda+h\left(X_{i},\bar{X}_{p^{*}\left(i\right)},\overline{X}_{p(i)}\right)+A_{i}.
\end{equation}
Condition \eqref{eq: conditional_moment_restriction} implies that  $A_i$ is conditionally mean zero given $X_i$, $\bar{X}_{p^{*}\left(i\right)}$ and $W^{*}_{i}$. 

Summarizing, the experimentally-induced restrictions \eqref{met: MET_implication_1}, \eqref{met: MET_implication_2}, and \eqref{met: MET_implication_3}, and our Assumptions \ref{ass:key_ass_1} and \ref{ass:key_ass_2} together imply that:
\begin{equation} \label{eq: estimation_moment}
    \mathbb{E}\left[\left.A_i\right|X_{i},\bar{X}_{p^{*}\left(i\right)},W^{*}_{i}\right]=0.
\end{equation}

The estimation simplifies if we impose a restriction on the peer attrition/non-compliance that is similar to Assumption \ref{ass:key_ass_2}, but is on the observable peer average: 
\begin{equation}
	\mathbb{E}\left[\left.\bar{X}_{p(i)}-\bar{X}_{p^{*}(i)}\right|X_{i},\bar{X}_{p^{*}\left(i\right)},W_{i}^{*}\right]=\mathbb{E}\left[\left.\bar{X}_{p(i)}-\bar{X}_{p^{*}(i)}\right|X_{i},\bar{X}_{p^{*}\left(i\right)}\right].\label{eq: key_ass_2 obs}
	\end{equation}
This restriction is directly testable. By equation \eqref{eq: key_ass_2 obs},
\[
\mathbb{E}\left[\left.\bar{X}_{p(i)}\right|X_{i},\bar{X}_{p^{*}\left(i\right)},W_{i}^{*}\right]=\mathbb{E}\left[\left.\bar{X}_{p(i)}\right|X_{i},\bar{X}_{p^{*}\left(i\right)}\right].
\]
If this restriction holds the outcome equation is as in \eqref{eq: plm_endogenous_version}, but with a redefined nonparametric $h$ that is a function of $X_{i}$ and $\bar{X}_{p^{*}\left(i\right)}$ only,
\begin{equation}
\begin{split}
    h\left(X_{i},\overline{X}_{p^{*}\left(i\right)} \right)\overset{def}{\equiv} & \alpha+X_{i}'\beta+g_{1}\left(X_{i},\overline{X}_{p^{*}\left(i\right)}\right)  + \mathbb{E}\left[\left.\bar{X}_{p(i)}\right|X_{i},\bar{X}_{p^{*}\left(i\right)}\right]' \gamma + \\ 
     & \rho g_{2}^*\left(X_{i},\overline{X}_{p^{*}\left(i\right)}\right)+ \left(X_{i}\otimes\mathbb{E}\left[\left.\bar{X}_{p(i)}\right|X_{i},\bar{X}_{p^{*}\left(i\right)}\right]\right )'\zeta,
\label{eq: h_redefined}    
\end{split}
\end{equation}

\noindent and $A_{i}$ is

\begin{equation}
\begin{split}
A_{i} \stackrel{def}{\equiv} & \left(V_{i}-g_{1}\left(X_{i},\overline{X}_{p^{*}\left(i\right)}\right)\right)+\rho\left(\overline {V}_{p\left(i\right)}-g^{*}_{2}\left(X_{i},\overline{X}_{p^{*}\left(i\right)}\right)\right) + U_{i}+\\
 & \left (\bar{X}_{p(i)}-\mathbb{E}\left[\left.\bar{X}_{p(i)}\right|X_{i},\bar{X}_{p^{*}\left(i\right)}\right] \right )' \gamma + \left (X_i  \otimes \left (\bar{X}_{p(i)}-\mathbb{E}\left[\left.\bar{X}_{p(i)}\right|X_{i},\bar{X}_{p^{*}\left(i\right)}\right] \right ) \right )' \zeta.
\label{eq: A_redefined}
\end{split}
\end{equation}

The conditional moment restriction in equation \eqref{eq: estimation_moment} also holds for this error.

Equations \eqref{eq: plm_endogenous_version} and \eqref{eq: estimation_moment} jointly define a partially linear model with an endogenous parametric component. This is a well-studied semiparametric model (see, for example, \citealp{Chen_et_al_EM2003}). The parameters $\delta$, $\eta$, and $\lambda$ are identified; $h\left(X_{i},\bar{X}_{p^{*}\left(i\right)}\right)$ is a nonparametric nuisance function.

We implement the partial linear IV estimator using the following approximation for $h(x,\bar{x})$:
\[
h\left(X_{i},\bar{X}_{p^{*}\left(i\right)}\right)\approx X_{i}'b+\bar{X}_{p^{*}\left(i\right)}'d+\left(X_{i}\otimes\bar{X}_{p^{*}\left(i\right)}\right)'f.
\]
For this approximation we estimate $\delta$, $\eta$, and $\lambda$ by linear IV fit of $Y_{i}$ onto a constant, $X_{i}$,
$\bar{X}_{p^{*}\left(i\right)},$ $\left(X_{i}\otimes\bar{X}_{p^{*}\left(i\right)}\right)$,
$W_{i}$, $\left(X_{i}\otimes W_{i}\right)$, and $\left(W_{i}\otimes\bar{X}_{p\left(i\right)}\right)$
using the excluded instruments $W_{i}^{*},\left(X_{i}\otimes W_{i}^{*}\right)$, and $\left(W_{i}^{*}\otimes\bar{X}_{p^{*}\left(i\right)}\right)$. Note that both assigned and realized peer groups enter the main equation.

As in the case with perfect compliance, we do not identify $\beta$, $\gamma$, and $\zeta$, again reflecting the inherent inability of the MET experiment to tell us anything about peer group effects. Nevertheless knowledge of $\delta$, $\eta$, and $\lambda$ is sufficient to identify the class of reallocation effects we focus upon.

\section{Data and tests of identifying assumptions}
\label{sec:data}

\subsection*{The Measures of Effective Teaching (MET) study}

The MET study was conducted during the 2009/10 and 2010/11 school years in elementary, middle, and high schools located in six large urban school districts in the United States.\footnote{The school districts are: Charlotte-Mecklenburg (North Carolina), Dallas Independent School District (Texas), Denver Public Schools (Colorado), Hillsborough County Public Schools (Florida), Memphis City Schools (Tennessee), and the New York City Department of Education (New York). The participation of school districts and schools in the MET study was voluntary.} The goal of the study was to examine determinants and consequences of teacher quality and teaching practices. In the first year of the study, the MET researchers collected detailed background information on teaching practices for each teacher as well as background performance measures for each student. In the second year of the study, MET researchers conducted a field experiment, randomly assigning teachers to classrooms within school-by-grade-by-subject cells (``randomization blocks''). The randomization blocks typically consisted of two to three classrooms each. Classroom composition was not manipulated as part of the study. For more details about the study design, see~\citet{WhiteRowan2018} and~\citet{Kaneetal2013}.

The dataset contains detailed information on the students, their classroom teachers, and their classroom peers. As our key measure of teaching practices, we use Danielson's \citeyearpar{Danielson2011} ``Framework for Teaching'' (FFT) measure, collected at baseline (school year 2009/10). The FFT is designed to provide feedback to teachers, but has also been used to study the impact of teaching practices on student outcomes~(c.f., \citealp{GarrettSteinberg2015,Aucejo2019}). The FFT is an observational measure: teachers are video-taped several times during a school year and subsequently rated \textendash{} using a specially designed rubric \textendash{} by trained raters (often former teachers).  

For our main outcome measure we use students' 2010/11 end-of-year standardized state test scores in the subjects mathematics and English language arts (ELA). The dataset also includes background characteristics from school district records for students (age, gender, race/ethnicity, special-education status, free/reduced-price lunch eligibility, gifted status, and whether a student is an English language learner) and for teachers (education, teaching experience in the district). Through a section identifier, we can match a student to her classroom peers. 

\subsection*{Sample and summary statistics}

In constructing our estimation sample, we closely follow~\citet{GarrettSteinberg2015}, who investigate the impact of FFT on students' test score outcomes in the MET study. We restrict our sample to all elementary- and middle-school students (grades 4--8) who took part in the randomization. Furthermore, we only include students with non-missing information on baseline and final test score outcomes as well as students with non-missing information on the characteristics of the assigned and realized teacher, and the characteristics of the assigned and realized classroom peers. Details of the data construction from the MET data files are provided in Appendix~{\ref{sec:data_appendix}}.

Our final sample consists of about 8,500 students and 614 teachers in math and of about 9,600 students and 649 teachers in ELA.\footnote{Appendix Table~\ref{tab:descriptives} summarizes all student and teacher variables.} The majority of the students (60 percent) are elementary school students (grades 4--5), and the remaining students are middle school students (grades 6--8). Nearly 60 percent of the students are economically disadvantaged, that is, eligible for free/reduced-price lunch.\footnote{This variable is missing for 30 percent of the students in the math sample and for 25 percent of students in the ELA sample.} The student sample is diverse, with equal proportions of white, black, and Hispanic students. 

Student test scores are centered at the relevant district-level mean and standardized by the relevant district-level standard deviation. MET students exceed their district average test score by about 0.15 standard deviations on average. The majority of the teachers are female (about 85 percent), and about two-thirds of them are white. Most of the teachers have substantial teaching experience \textendash{} seven years on average \textendash{} and about 40 percent of the teachers have graduated from a  Master's program.\footnote{Teaching experience and teacher education are missing for about 30 percent of teachers.} 

\subsubsection*{Teaching practices}

The MET data includes ratings of teaching practices for two domains of the FFT, ``Classroom Environment'' and ``Instruction.'' These domains are divided into four components each,\footnote{The components of the domain ``Classroom Environment'' are: creating an environment of respect and rapport; establishing a culture for learning; managing classroom procedures; managing student behavior. The components of the domain ``Instruction'' are: communicating with students; using questioning and discussion techniques; engaging students in learning; using assessment in instruction.} and each component is rated on a four-point scale by each rater (unsatisfactory, basic, proficient, distinguished), with high scores in a category indicating that a teacher is closer to an ideal teaching practice according to the FFT. The teachers are video-taped at least twice during the school year during lessons, and each video is rated independently by at least two raters. In our analysis, we average the scores across all videos, raters, components, and domains. This aggregation of the scores is the most common one in the literature~(cf., \citealp{GarrettSteinberg2015,Aucejo2019}). 

The FFT is on average 2.5 in math and 2.6 in ELA, which corresponds to a rating between ``basic'' and ``proficient''; for the purpose of our analysis, we create three categories of FFT (see Appendix Figure~\ref{fig:baseline-fft}). We set the cutoffs at FFTs of 2.25 and 2.75. In our sample, low-FFT teachers have ``basic'' teaching practices on average (average FFT of 2.1 in both math and ELA), and high-FFT teachers have ``proficient'' teaching practices on average (average FFT of 2.9 in both math and ELA); 18 percent of the teachers in math and 14 percent of the teachers in ELA are classified as having a low FFT, 62 percent of the teachers in math and 58 percent of the teachers in ELA are classified as having a middle FFT, and 20 percent of the teachers in math and 28 percent of the teachers in ELA are classified as having a high FFT. Our results are not sensitive to the exact position of the cutoffs, and we investigate a model with two or four, instead of three, FFT categories in our sensitivity checks below. This categorization also creates variation in FFT levels within randomization blocks: in our sample, 65 percent of randomization blocks in math, and 70 percent of randomization blocks in ELA include teachers with different levels of FFT (for example, both a low- and a middle-FFT teacher). 

\subsubsection*{Test score outcomes}

The end-of-year test scores provided in the MET data are $z$-scores, i.e., they are standardized such that district-wide, they will be mean zero with unit standard deviation. Since MET schools are not representative at the district level, the mean test scores may \textendash{} and do \textendash{} deviate from zero. We use the 2010/11 $z$-score as our outcome variable.

To identify reallocation effects, we split 2009/10 baseline test scores into three bins, corresponding to terciles of the within-district $z$-score distribution. In our estimation sample, 27 percent of the students in math, and 26 percent in ELA, have ``low'' baseline test scores, 36 percent of the students in math, and 35 percent of the students in ELA, have ``middle'' baseline test scores, and 36 percent of the students in math, and 39 percent of the students in ELA, have ``high'' baseline test scores.\footnote{These numbers are not exactly 33 percent in each category, because the students in our sample have on average higher baseline test scores than the full student population in a district.} To include classroom peers into the analysis, we compute the fraction of each student's classmates with high, middle, and low baseline test scores (leave-own-out means).

\subsubsection*{Non-compliance}

As outlined in Section~\ref{sec:model}, some teachers and students switched classrooms or schools before the start of the school year, which we take into account in our identification strategy. In our sample, 69 percent of the students in math, and 73 percent of the students in ELA, are actually taught by their randomly assigned teachers. This level of non-compliance is high enough to make analyses which ignore it potentially problematic.

We also observe changes in classroom peers after the randomization but before the beginning of the school year. Changes in assigned peers were driven by students who leave schools, repeat a grade or, in some cases, by the need to adjust their schedule. On average, students' realized peers, however, are not appreciably different from their assigned ones. The difference in baseline $z$-scores between the assigned and the realized peers amounts to just 0.02 standard deviations on average both in math and in ELA.

\subsection*{Tests of identifying assumptions and restrictions}

In this section, we report the results of the series of specification tests discussed earlier; specifically tests designed to assess the plausibility of Assumptions \ref{ass:key_ass_1} and \ref{ass:key_ass_2}. These two assumptions impose restrictions on the nature of non-compliance by students and teachers. We further \emph{directly} test restriction \eqref{eq: key_ass_2 obs}, which is a restriction on the nature of non-compliance by peers. We also assess the quality of the initial MET randomization of teachers to classrooms.

To test whether the randomization was successful in balancing student characteristics across teachers with different levels of FFT we regress the FFT of a student's assigned teacher on the student's own characteristics, controlling for randomization block fixed effects. None of the student characteristics predict assigned teacher's FFT, individually or jointly, which confirms that the randomization indeed ``worked'' (see Appendix Table~\ref{tab:balancing}).

Covariate balance, however, is not a sufficient condition to identify reallocation effects under non-compliance by both students and teachers (see Section~\ref{sec:model}). Assumption \ref{ass:key_ass_1} requires that own and assigned peer and teacher observables should not predict the difference between realized and assigned unobserved teacher quality. Since this assumption involves a statement about unobserved variables, we cannot test it directly. Instead we ``test" it indirectly as described in Section~\ref{sec:model}. Specifically we use those teacher background characteristics that are not part of the the model as replacements for the unobserved quality of a teacher: a teacher's demographics, experience, and education. We regress the difference between realized and assigned teacher characteristics on the student's baseline test score, the FFT of the assigned teacher, and the average baseline test score of the assigned peers. Consistent with Assumption \ref{ass:key_ass_1}, these variables do not jointly predict differences between the characteristics of the assigned and realized teacher in any of the regression fits (see Appendix Table~\ref{tab:assumption-1}). 

Assumption \ref{ass:key_ass_2} states that differences between realized and assigned unobserved peer quality cannot be predicted by assigned teacher observables, conditional on own baseline achievement and the baseline achievement of assigned peers. Again, we can only perform an indirect test of this assumption. To do so we regress differences between the assigned and realized characteristics of classroom peers onto the FFT of the assigned teacher, controlling for own baseline test scores and assigned peers' average baseline test scores. We do not find, consistent with Assumption \ref{ass:key_ass_2}, that teacher FFT predicts differences between the characteristics of the assigned and realized peers (see Appendix Tables~\ref{tab:assumption-2_math} and~\ref{tab:assumption-2_ela}).

Finally, we directly assess restriction \eqref{eq: key_ass_2 obs}, which implies that the baseline test scores of realized peers cannot be predicted by assigned teacher FFT. We test this restriction by regressing the baseline test scores of realized peers onto a student's baseline test score, the baseline test scores of her assigned peers, as well as the FFT of her assigned teacher. We find that this restriction also holds in our data~(see Appendix Table~\ref{tab:assumption-3}). 

\section{Computing reallocation effects}

To construct estimates of the average effect of reassigning teachers across classrooms we proceed in three steps. First, as outlined in Section \ref{sec:model}, we estimate (a subset of) the parameters of the educational production function. Second, we feed these parameters into a linear programming problem to compute an optimal assignment (i.e., an assignment that maximizes aggregate achievement). Third, we compare the aggregate outcome under the optimal assignment to the aggregate outcome under the status quo assignment, as well as to that under the worst assignment (i.e., the one which minimizes aggregate achievement). These comparisons provide a sense of the magnitude of achievement gains available from improved teacher assignment policies.

\subsection*{Estimation of the education production function}

The specification that we use to estimate the parameters of the educational production function coincides with equation \eqref{eq: plm_endogenous_version}. Since randomization was carried out within randomization blocks, we additionally include randomization block fixed effects in this regression model. We then estimate the model's parameters by the method of instrumental variables (IV) as described in  Section \ref{sec:model}. Tests for instrument relevance following~\citet{Sanderson2016} confirm that our IV estimates are unlikely to suffer from weak-instrument bias (see Appendix Table~\ref{tab:firststage}).

\subsection*{Computing and characterizing an optimal assignment}
\label{sec:optimal}

The aim of the analysis is to find an assignment of teachers to classrooms that improves student outcomes. What is meant by an ``optimal" assignment depends, of course, on the objective function.

In this paper we choose to maximize aggregate outcomes (i.e., the sum of all students' test scores). This is the ``simplest'' objective we can consider, it is straightforward to compute, justifiable from a utilitarian perspective, and easy to interpret. We would like to emphasize, however, that our analysis can be modified to accommodate other objective functions \textendash{} in some cases easily, in others with more difficulty. In practice, for example, the social planner may care about both the aggregate outcome as well as inequality, especially across identifiable subgroups. 

Our intention is not to advocate for maximization of the aggregate outcome in practice, although doing so is appropriate in some circumstances; rather this objective provides a convenient starting point and makes our analysis comparable to that of other educational policy evaluations (e.g., the typical regression-based class-size analysis provides an estimate of the effect of class-size on \emph{average} achievement).

To determine the optimal allocation, we feed the \textit{estimated} parameters from equation (\ref{eq: plm_endogenous_version}), specifically $\hat{\delta}$, $\hat{\eta}$, and $\hat{\lambda}$, into a linear program. Given these parameters, we can compute, for each student, her predicted outcome when taught by a low-, middle-, or high-FFT teacher, leaving the original classroom composition unchanged. 

Formally, for each student $i$, we compute three counterfactual outcomes  $\widehat{Y}_{i}(w)$, $w \in \{w_L, w_M, w_H\}$, corresponding to an assignment to a low-, middle- or high-FFT teacher. Aggregated to the classroom level this yields, in an abuse of notation, three counterfactual classroom-level test score aggregates, $\widehat{{Y}}_{c}(w)=\sum_{i\in s\left(c\right)}\widehat{Y}_{i}(w)$, with $w \in \{w_L, w_M, w_H\}$ (here $s\left(c\right)$ denotes the set of indices for students in classroom $c$).

By aggregating the counterfactual outcomes to the \textit{classroom} level, we  transform the assignment problem from a many-to-one matching problem to a one-to-one matching problem. This approach is only suitable because the configuration of students across classrooms remains fixed; we only consider the effects of reassigning teachers across existing classrooms of students.\footnote{As noted earlier the effects of classroom composition changes on student achievement are unidentifiable here in any case.} The one-to-one matching problem is a special linear program, a transportation problem, which is easily solvable.\footnote{Appendix~\ref{sec:optimal-compute} contains~the details on how we specify the transportation problem. To compute the optimal assignment, we use the function \tt{lpSolve} in $R$.} 

There are a few additional constraints we impose to make the reallocation exercise realistic. First, we do not allow teachers to be reassigned across districts or across school types (i.e., elementary or middle school). We also present, as a sensitivity check, a version of the allocation where we only allow teachers to be reassigned within their randomization block. Second, there are a few teachers that teach several sections of a class. In this case, we treat these sections as clusters, and allocate one teacher to all sections in each such cluster.

When computing counterfactual outcomes for each student, we omit the nonparametric component, $h(x,\bar{x})$, as well as the randomization block fixed effects and the main effects of own baseline achievement. This is appropriate because we report the difference between the sum of aggregate outcomes for two allocations where the nonparametric component, the fixed effects, and the main effects of own baseline achievement cancel out.

In addition to the optimal assignment, we also compute a ``worst'' possible assignment, i.e., the assignment that minimizes aggregate test scores. The difference between the aggregate outcome for the best and worst assignment is the maximal reallocation gain. This provides an upper bound on the magnitude of student achievement gains that teacher reassignments might yield in practice.

\subsection*{Average Reallocation Effects}

An individual  reallocation effect, or reallocation gain, is defined as the difference between an individual student's outcome under two  assignments. For example, one can compute, for each student, $\widehat{Y}_{i}(\tilde{W}_{i}^{opt})$, i.e., the predicted outcome under the optimal allocation, where $\tilde{W}_{i}^{opt}$ takes the values $w_L$, $w_M$, or $w_H$, depending on whether the student would be assigned to a low-, middle-, or high-FFT teacher in an optimal allocation. Similarly, one can compute the same parameter for each student under the status quo, which we denote as $\widehat{Y}_{i}(W_{i})$. In MET schools the status quo assignment was induced by random assignment of teachers to classrooms within randomization blocks. The allocation of teachers across schools within a given district was, of course, non-random, and possibly non-optimal from the standpoint of maximizing the aggregate outcome. An individual reallocation gain can then be computed as the difference between these two outcomes, $\widehat{Y}_{i}(\tilde{W}_{i}^{opt})-\widehat{Y}_{i}(W_{i})$. The optimal assignment is not the assignment that maximizes the predicted outcome for student $i$, but the assignment that maximizes the aggregate outcome across all the classrooms in the sample (subject to the various constraints we impose on the problem \textendash{} like ruling out reassignments across school districts). 

These individual gains can be aggregated in many ways in order to create policy-relevant parameters. We start by defining our key parameter of interest, the \textit{average reallocation effect}, as

\begin{equation}
\widehat{ARE} = \frac{1}{N}\sum_{i = 1}^N\left[\widehat{Y}_{i}(\tilde{W}_{i}^{opt})-\widehat{Y}_{i}(W_{i})\right].
\end{equation}

We also consider conditional average reallocation effects, that is, average reallocation effects for students with varying baseline characteristic $x$ (e.g., students with low, middle, and high baseline test scores): 

\begin{equation}
\widehat{ARE} (x) = \frac{1}{\sum_{i = 1}^{N} \mathbf{1}(X_i=x)} \sum_{i = 1}^{N} \mathbf{1}(X_i=x)\left[\widehat{Y}_{i}(\tilde{W}_{i}^{opt})-\widehat{Y}_{i}(W_{i})\right].
\end{equation}

Below we show that only about one-half of MET students would experience a teacher change when switching from the status quo teacher assignment to an optimal one. The reallocation effect for students in classrooms not assigned a different teacher is, of course, zero. This motivates a focus on the average achievement gain among those students who \emph{are} assigned a different teacher. We define the reallocation effect \textit{conditional on being reassigned} as:

\begin{multline}
\widehat{AREC} (x) = \frac{1}{\sum_{i = 1}^{N} \mathbf{1}(X_i=x) \mathbf{1}(W_{i}\neq\tilde{W}_{i}^{opt})} \\
\times \sum_{i = 1}^{N} \mathbf{1}(X_i=x)\mathbf{1}(W_{i}\neq\tilde{W}_{i}^{opt}) \left[\widehat{Y}_{i}(\tilde{W}_{i}^{opt})-\widehat{Y}_{i}(W_{i})\right].
\end{multline}

We can, of course, not condition on $x$ as well.

Finally, the maximal reallocation gain is obtained if we compare the average outcome under an optimal assignment with that under a worst assignment

\begin{equation}
\widehat{ARE}^{max} (x) = \frac{1}{\sum_{i = 1}^{N} \mathbf{1}(X_i=x)} \sum_{i = 1}^{N} \mathbf{1}(X_i=x)\left[\widehat{Y}_{i}(\tilde{W}_{i}^{opt})-\widehat{Y}_{i}(\tilde{W}_{i}^{worst})\right],
\end{equation}

\noindent where $\widehat{Y}_{i}(\tilde{W}_{i}^{worst})$ is the predicted outcome for student $i$ under an allocation that minimizes aggregate test scores.

\subsection*{Inference on reallocation effects}
\label{sec:inference}

We use the Bayesian bootstrap to quantify our (posterior) uncertainty about AREs. We treat each teacher-classroom pair as an i.i.d. draw from some unknown (population) distribution. Following \cite{Chamberlain_Imbens_JBES03}, we approximate this unknown population by a multinomial, to which we assign an improper Dirichlet prior. This leads to a posterior distribution which (i) is also Dirichlet and (ii) conveniently only places probability mass on data points observed in our sample. 

Mechanically, we draw $C$ standard exponential random variables and weigh each student in section/classroom $c=1, \dots, C$ with the $c^{th}$ weight (i.e., all students in the same section have the same weight). Using this weighted sample, we compute the IV regression fit, solve for the optimal (worst) assignment, and compute the various reallocation effects. We repeat this procedure 1,000 times. This generates 1,000 independent draws from the posterior distribution of the ARE.

Formally this approach to inference is Bayesian. Consequently the ``standard errors" we present for our ARE estimates summarize dispersion in the relevant posterior distribution (not variability across repeated samples).\footnote{The ``standard errors" for the AREs are constructed as follows. Let $\hat{\theta}^{(b)}$ be the estimate of the reallocation effect in $b^{th}$ the bootstrap sample (or equivalently the $b^{th}$ draw from the posterior distribution for $\theta$) and $\hat{\theta}$ the reallocation effect in the original sample. Consider the centered statistic $\hat{\theta}^{(b)} - \hat{\theta}$; let $q(0.025)$ and $q(0.975)$ be the 0.025 and 0.975 quantiles of its bootstrap/posterior distribution. We then construct the interval $[\hat{\theta} - q(0.975), \hat{\theta} - q(0.025)]$ (see ~\citealp{Hansen:2018}). We compute the corresponding standard error by dividing this interval by $2\Phi^{-1}(0.975) \approx 3.92$, where $\Phi(.)$ is the standard normal CDF.} An alternative, frequentist, approach to inference is provided by~\citet{Hsieh2018}. They transform the problem of inference on the solution to a linear program into inference on a set of linear moment inequalities. If the binding constraints are the same over the bootstrap distribution, then inference based on the Bayesian bootstrap will be similar to that based on moment inequalities (see also \citealp{Graham_Imbens_Ridder_USC2007} and \citealp{Bhattacharya_JASA2009}).

\section{Results}

\subsection*{Regression results}


Before reporting parameter estimates for our preferred model \textendash{} equation \eqref{eq: plm_endogenous_version} above with three categories of FFT, own and peer baseline achievement each \textendash{}  we present those for a more conventional model with teacher FFT, student and peer average baseline test scores all entering linearly. For this specification we leave the FFT and baseline test scores undiscretized. These initial results replicate and expand upon the prior work of~\citet{GarrettSteinberg2015}, who study the impact of FFT on student achievement in (approximately) the same sample.\footnote{\citet{Aucejo2019} study the impact of FFT in the MET data, but focus on math classrooms in elementary schools.} 

These initial results indicate that teacher FFT does not affect students' test scores on average (see Table~\ref{tab:results_lin_iv}, Panel A). This is consistent with the findings of~\citet{GarrettSteinberg2015}. Table~\ref{tab:results_lin_iv} presents estimates, using the instrumental variables described in Section \ref{sec:model}. Here this involves using the FFT score of a student's randomly \emph{assigned} teacher as an instrument for that of her \emph{realized} teacher (Panels A--C), and the average baseline test score of her assigned peers as an instrument for the average baseline test score of her realized peers (Panel B).

\begin{table}[h!tbp]
	\centering
	\caption{IV regression results of the linear model. Dependent variables: student test score outcomes}
\begin{small}
\begin{tabular}{lrcccccc}
	\toprule
	\toprule
	&            & (1)        & (2)        & (3)        & (4)        & (5)        & (6) \\
	&            & \multicolumn{2}{c}{A. Only teacher} & \multicolumn{2}{c}{B. Full model} & \multicolumn{2}{c}{C. Without} \\
	&            & \multicolumn{2}{c}{effects} &            &            & \multicolumn{2}{c}{teacher $\times$ peer} \\
	&            & \multicolumn{2}{c}{}    &            &            & \multicolumn{2}{c}{interactions} \\
	&            & Math       & ELA        & Math       & ELA        & Math       & ELA \\
	\hline
	$\delta$ & \multicolumn{1}{l}{FFT } & 0.013      & 0.079      & -0.004     & 0.104      & -0.004     & 0.076 \\
	&            & (0.093)    & (0.076)    & (0.091)    & (0.075)    & (0.091)    & (0.075) \\
	$\eta$ & \multicolumn{1}{l}{FFT $\times$ baseline} &            &            & 0.096**    & 0.073*     & 0.098**    & 0.049 \\
	&            &            &            & (0.042)    & (0.039)    & (0.046)    & (0.040) \\
	$\lambda$ & \multicolumn{1}{l}{FFT $\times$ avg. peer baseline} &            &            & 0.016      & -0.172     &            &  \\
	&            &            &            & (0.103)    & (0.197)    &            &  \\
	$\beta$ & \multicolumn{1}{l}{Baseline } & 0.749***   & 0.693***   & 0.505***   & 0.512***   & 0.498***   & 0.565*** \\
	&            & (0.011)    & (0.011)    & (0.109)    & (0.099)    & (0.119)    & (0.103) \\
	\midrule
	$R^2$ &            & 0.700      & 0.632      & 0.700      & 0.621      & 0.700      & 0.632 \\
	N          &            & 8,534      & 9,641      & 8,534      & 9,641      & 8,534      & 9,641 \\
	\bottomrule
	\bottomrule
\end{tabular}%
\end{small}
\caption*{\footnotesize \textit{Note:} The dependent variables are subject-specific test score outcomes in math and ELA. The specifications include linear terms for FFT, individual and peer baseline test scores. The instrumental variables are based on assigned teacher FFT (Panels A--C) and assigned peer baseline test scores (Panel B). All regressions control for the $h(x,\bar{x})$ function (see Section~\ref{sec:model}) and for randomization block fixed effects. Analytic standard errors, clustered by randomization block, are in parentheses. \\*** significant at the 1\%-level ** significant at the 5\%-level * significant at the 10\%-level.}
\label{tab:results_lin_iv}
\end{table}

OLS estimates of the same model are reported in Appendix Table~\ref{tab:results_lin_ols}. The coefficient on FFT is statistically significant and positive in the OLS results. The discrepancy between the OLS and IV results may reflect the impact of correcting for non-compliance, as described above, or simply reflect the greater sampling variability of the IV estimates.

Next we add the interactions of teacher FFT with both the baseline student, and peer average, test scores. This provides an initial indication of whether any complementarity between teacher FFT and student baseline achievement is present. These results suggest that high-FFT teachers are more effective at teaching students with higher baseline scores (see Table~\ref{tab:results_lin_iv}, Panels B and C). This result is significant at the 5-percent level in the math sample, but only weakly significant in the ELA sample. The magnitudes and precision of the teacher-student match effects are similar across the IV and OLS estimates (the latter reported in Appendix Table~\ref{tab:results_lin_ols}). The coefficient on the interaction between teacher FFT and peer average baseline achievement (Panel B) is poorly determined (whether estimated by IV or OLS).

\begin{table}[h!tbp]
	\centering
	\caption{IV regression results of the 3 $\times$ 3 model. Dependent variables: student test score outcomes}
\begin{small}
\begin{tabular}{lrcccccc}
	\toprule
	\toprule
	&            & (1)        & (2)        & (3)        & (4)       & (5)       & (6) \\
	&            & \multicolumn{2}{c}{A. Only teacher} & \multicolumn{2}{c}{B. Full model} & \multicolumn{2}{c}{C. Without} \\
	&            & \multicolumn{2}{c}{effects} &            &            & \multicolumn{2}{c}{teacher $\times$ peer} \\
	&            & \multicolumn{2}{c}{}    &            &            & \multicolumn{2}{c}{interactions} \\
	&            & Math       & ELA        & Math       & ELA        & Math       & ELA \\
	\midrule
	$\delta$      & \multicolumn{1}{l}{FFT middle} & 0.069      & -0.029     & -0.145     & -0.219     & 0.027      & -0.082 \\
	&            & (0.053)    & (0.048)    & (0.138)    & (0.173)    & (0.060)    & (0.059) \\
	& \multicolumn{1}{l}{FFT high} & 0.038      & -0.058     & -0.547     & -0.137     & -0.155     & -0.154* \\
	&            & (0.067)    & (0.065)    & (0.380)    & (0.203)    & (0.103)    & (0.083) \\
	$\eta$        & \multicolumn{1}{l}{FFT middle } &            &            &            &            &            &  \\
	& \multicolumn{1}{l}{\hspace{0.1cm} $\times$  baseline middle} &            &            & 0.040      & 0.057      & 0.052      & 0.067 \\
	&            &            &            & (0.060)    & (0.067)    & (0.060)    & (0.063) \\
	& \multicolumn{1}{l}{\hspace{0.1cm} $\times$ baseline high} &            &            & 0.015      & 0.076      & 0.050      & 0.097 \\
	&            &            &            & (0.084)    & (0.081)    & (0.076)    & (0.082) \\
	& \multicolumn{1}{l}{FFT high} &            &            &            &            &            &  \\
	& \multicolumn{1}{l}{\hspace{0.1cm} $\times$ baseline middle} &            &            & 0.184**    & 0.150**    & 0.196**    & 0.127* \\
	&            &            &            & (0.082)    & (0.074)    & (0.084)    & (0.072) \\
	& \multicolumn{1}{l}{\hspace{0.1cm} $\times$ baseline high} &            &            & 0.226**    & 0.187**    & 0.265**    & 0.149 \\
	&            &            &            & (0.099)    & (0.092)    & (0.100)    & (0.095) \\
	$\lambda$     & \multicolumn{1}{l}{FFT middle} &            &            &            &            &            &  \\
	& \multicolumn{1}{l}{\hspace{0.1cm} $\times$ fraction peers middle} &            &            & 0.318      & 0.288      &            &  \\
	&            &            &            & (0.299)    & (0.401)    &            &  \\
	& \multicolumn{1}{l}{\hspace{0.1cm} $\times$ fraction peers high} &            &            & 0.239      & 0.161      &            &  \\
	&            &            &            & (0.205)    & (0.288)    &            &  \\
	& \multicolumn{1}{l}{FFT high} &            &            &            &            &            &  \\
	& \multicolumn{1}{l}{\hspace{0.1cm} $\times$ fraction peers middle} &            &            & 0.641      & 0.227      &            &  \\
	&            &            &            & (0.513)    & (0.389)    &            &  \\
	& \multicolumn{1}{l}{\hspace{0.1cm} $\times$ fraction peers high} &            &            & 0.460      & -0.355     &            &  \\
	&            &            &            & (0.409)    & (0.343)    &            &  \\
	&            &            &            &            &            &            &  \\
	$\beta$       & \multicolumn{1}{l}{Baseline middle} & 0.888***   & 0.739***   & 0.843***   & 0.699***   & 0.840***   & 0.682*** \\
	&            & (0.077)    & (0.084)    & (0.092)    & (0.096)    & (0.092)    & (0.094) \\
	& \multicolumn{1}{l}{Baseline high} & 1.622***   & 1.555***   & 1.578***   & 1.503***   & 1.550***   & 1.470*** \\
	&            & (0.103)    & (0.105)    & (0.124)    & (0.118)    & (0.119)    & (0.115) \\
	\midrule
	$R^2$       &            & 0.617      & 0.554      & 0.616      & 0.552      & 0.617      & 0.555 \\
	N          &            & 8,534      & 9,641      & 8,534      & 9,641      & 8,534      & 9,641 \\
	\bottomrule
	\bottomrule
\end{tabular}%
\end{small}
	\caption*{\footnotesize \textit{Note:} The dependent variables are subject-specific test score outcomes in math and ELA. The instrumental variables are based on assigned teacher FFT (Panels A--C) and assigned peer baseline test scores (Panel B). All regressions control for the $h(x,\bar{x})$ function (see Section~\ref{sec:model}) and for randomization block fixed effects. Analytic standard errors, clustered by randomization block, are in parentheses. \\ *** significant at the 1\%-level ** significant at the 5\%-level * significant at the 10\%-level.}
	\label{tab:results_3-by-3_iv}
\end{table}

In Table~\ref{tab:results_3-by-3_iv} we report IV estimates of our preferred $3 \times 3$ model. This corresponds to equation \eqref{eq: plm_endogenous_version} with both $W_i$ and $X_i$ consisting of dummy variables for middle and high FFT and baseline test scores respectively. In this specification, the teacher FFT main effects remain insignificant. We do, however, find positive match effects between teacher FFT and student baseline scores. These are significant for the high-FFT teachers (i.e., teachers with a ``proficient'' score on average). Students with middle or high baseline test score levels score significantly higher on end-of-year achievement tests when matched with a high-FFT teacher, compared to students with low baseline test scores (see Table~\ref{tab:results_3-by-3_iv}, Panels B and C).\footnote{In contrast, the coefficients on interactions of the teacher FFT dummy variables with the peer composition variables are imprecisely determined.}  The OLS estimation results, reported in the appendix, are qualitatively similar to the IV ones; although the interactions of the middle-FFT dummy variable with the middle- and high-baseline student test score dummies are also significant when estimated by OLS~(see Appendix Table~\ref{tab:results_3-by-3_ols}).

To compute optimal assignments  and average reallocation effects, we use the IV estimates presented in columns 5--6 of Table~\ref{tab:results_3-by-3_iv}. These specifications omit the FFT-by-peer composition interaction terms, whose coefficients are poorly determined in all specifications (whether fitted by IV or OLS). Omitting these terms has little effect on either the location or the precision of the coefficients on the FFT-by-baseline interactions. In Appendix Tables \ref{tab:teach-by-peer_math} and \ref{tab:teach-by-peer_ela} we also present average rellocation effects based upon the specifications in columns 3--4 of Table~\ref{tab:results_3-by-3_iv}. These ARE estimates are larger, albeit less precisely determined. 

\subsection*{Characterization of an optimal reassignment}

If an optimal assignment is similar to the status quo one, then large reallocation effects are unlikely. Therefore, before quantifying any reallocation effects, we first discuss how an optimal assignment and a worst assignment differ from the status quo. 

Our primary reassignment policy considers reassignments of teachers across classrooms district-wide (that is, we allow teachers to move to a different school within their district). We do restrict teachers to teach at the same level (e.g., elementary school teachers do not move to middle school classrooms). Under this scenario we find that, when moving from the status quo to an optimal assignment, 49 percent of the students in the math sample, and 47 percent of the students in the ELA sample, are assigned to a new teacher. The balance of students remain with their status quo teacher.

\begin{figure}[h!tbp]
	\centering
		\caption{Assortativeness of the optimal allocation in comparison with the status quo}
	\includegraphics[width=0.8\linewidth]{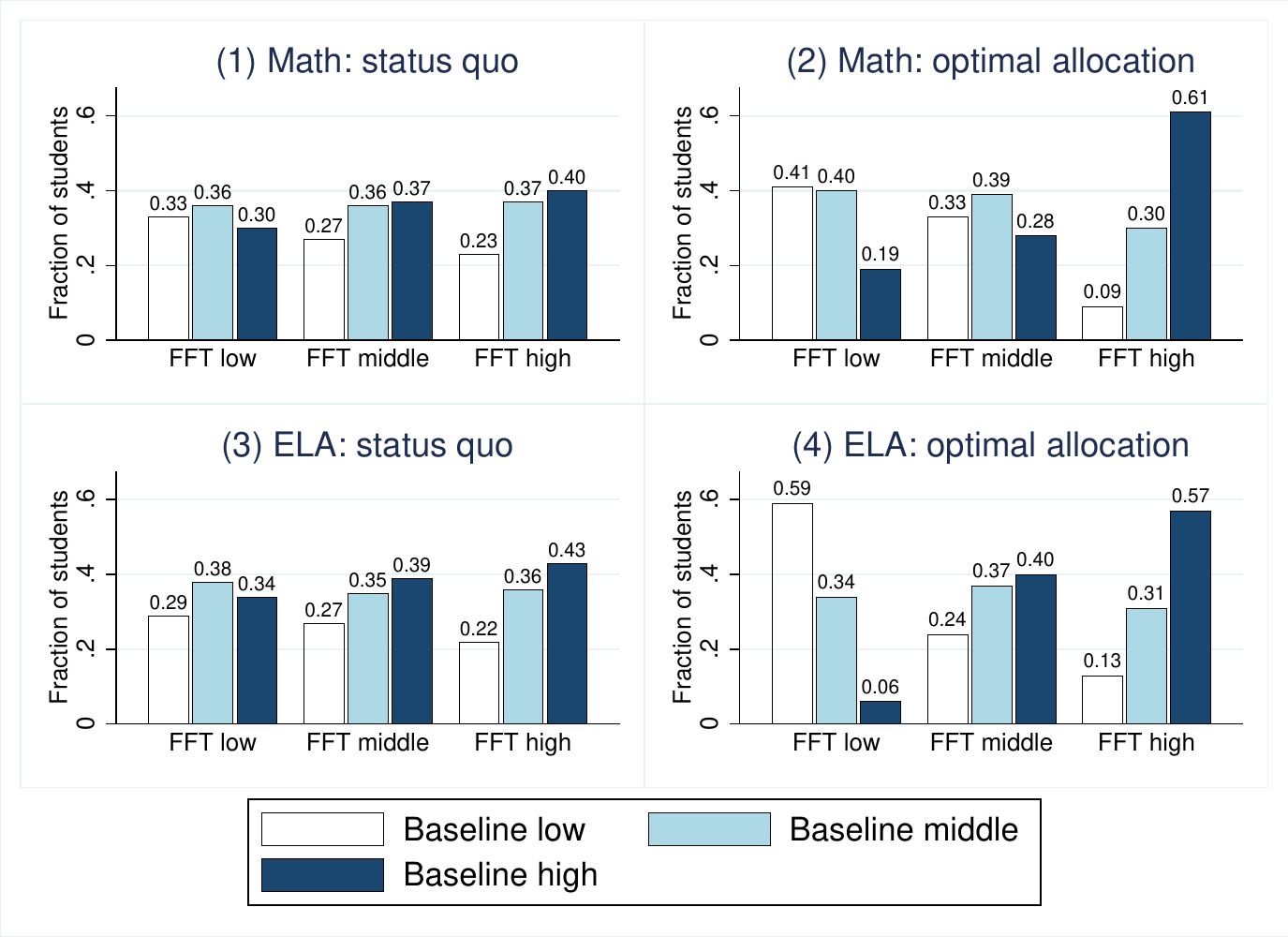}
		\caption*{\footnotesize \textit{Note:} The figure compares the status quo allocation (Panels 1 and 3) with the optimal allocation (Panels 2 and 4). The bars represent the fractions of students with low, middle, and high baseline test scores that a teacher with low, middle, and high FFT is assigned to on average under each of the allocations. For each teacher type, the fractions add up to 1. The optimization is carried out within school types (elementary or middle school) and districts.}
	\label{fig:assortativeness}
\end{figure}

It is interesting to examine how the reallocation changes the ``assortativeness'' of the assignment. An assignment is characterized as positive assortative if students with higher baseline test scores are more frequently matched with higher-FFT teachers, compared to a random assignment. Indeed, we observe this in our data for the optimal assignment. Figure~\ref{fig:assortativeness} displays, for each level of FFT, the distribution of student baseline test scores in the average class a teacher of that FFT level  is assigned to. In both the math and in the ELA sample, the optimal allocation is more assortative than the status quo. In the status quo in math, for instance, a high-FFT teacher has on average 23 percent of students with low baseline test scores and 40 percent of students with high baseline test scores. In the optimal allocation, the fraction of students with low baseline test scores drops to 9 percent, and the fraction of students with high baseline test scores increases to 61 percent on average. The optimal allocation in the ELA sample displays a similar pattern. The positive assortativeness arises in the optimal allocation because students with high baseline test scores benefit more from  a high-FFT teacher, compared to students with low baseline test scores (see Table~\ref{tab:results_3-by-3_iv}). The worst allocation displays negative assortativeness (see Appendix Figure~\ref{fig:assortativeness-worst}).

\subsection*{Average reallocation effects}

Tables~\ref{tab:reallocation-results_math} and~\ref{tab:reallocation-results_ela} present ARE estimates. In Panel A an optimal assignment is compared with the status quo; while in Panel B optimal assignments are compared with ``worst" assignments. We presents results for all students (Panels A.I and B.I), as well as for just those students who experience a teacher change as part of the reallocation (Panels A.II and B.II).

In the math sample (Table~\ref{tab:reallocation-results_math}), the optimal allocation improves average test scores by 1.7 percent of a test score standard deviation compared to the status quo. This effect is precisely determined with a Bayesian bootstrap standard error of 0.6 percent. The reported effects are largely driven by students with high baseline test scores. These students gain 2.8 percent of a test score standard deviation on average; students with middle and low baseline test scores, in contrast, gain only 1.2 and 1.4 percent of a test score standard deviation, respectively (on average).

Since only half of the students experience a change in their teacher, the average effect represents an equal-weighted mixture of a zero effect and a positive effect on those students who do experience a change in teachers. The average effect for the latter group is 3.6 percent of a test score standard deviation (Panel A.II, SE = 1.2 percent). 

\begin{table}[h!tbp]
	\centering
	\caption{Average reallocation gains in math. Gains expressed in test score standard deviations}
\begin{small}
\begin{tabular}{lccccccccc}
	\toprule
	\toprule
	& (1)        & (2)        & (3)        & (4)        &            & (5)        & (6)        & (7)        & (8) \\
	\midrule
	& \multicolumn{9}{c}{Panel A. Optimal versus status quo} \\
	& \multicolumn{4}{c}{A.I Full sample}               &            & \multicolumn{4}{c}{A.II Conditional on being reallocated} \\
	\cmidrule{2-5}\cmidrule{7-10}           & all students & high       & middle     & low        &            & all students & high       & middle     & low \\
	Gain       & 0.017      & 0.028      & 0.012      & 0.014      &            & 0.036      & 0.059      & 0.028      & 0.026 \\
	SE         & (0.006)    & (0.014)    & (0.008)    & (0.011)    &            & (0.012)    & (0.029)    & (0.019)    & (0.019) \\
	N          & 8,534      & 2,332      & 3,108      & 3,094      &            & 4,107      & 1,121      & 1,380      & 1,606 \\
	&            &            &            &            &            &            &            &            &  \\
	& \multicolumn{9}{c}{Panel B. Optimal versus worst allocation} \\
	\cmidrule{2-10}           & \multicolumn{4}{c}{B.I Full sample}               &            & \multicolumn{4}{c}{B.II Conditional on being reallocated} \\
	\cmidrule{2-5}\cmidrule{7-10}           & all students & high       & middle     & low        &            & all students & high       & middle     & low \\
	Gain       & 0.040      & 0.072      & 0.019      & 0.038      &            & 0.060      & 0.103      & 0.032      & 0.053 \\
	SE         & (0.012)    & (0.034)    & (0.015)    & (0.026)    &            & (0.019)    & (0.048)    & (0.022)    & (0.037) \\
	N          & 8,534      & 2,332      & 3,108      & 3,094      &            & 5,746      & 1,626      & 1,912      & 2,208 \\
	\bottomrule
	\bottomrule
\end{tabular}%
\end{small}
	\caption*{\footnotesize \textit{Note:} The table shows the average reallocation gains from implementing the optimal assignment instead of the random assignment (status quo) in Panel A, and the average reallocation gains from implementing the optimal assignment instead of the worst assignment in Panel B. The gains are expressed in test score standard deviations. The computations are based on a 3 $\times$ 3 model without teacher-by-peer interactions (see Table~\ref{tab:results_3-by-3_iv}, column 5). The results are presented separately for the full sample of students (columns 1--4), and for the sample of classrooms that would get a new teacher as a result of the reallocation (columns 5--8). The reassignments are carried out within school types and districts. Standard errors are in parentheses and computed using the Bayesian bootstrap with 1,000 replications (see Section~\ref{sec:inference}). High/middle/low: students in the top/middle/bottom tercile of the baseline test score distribution.}
	\label{tab:reallocation-results_math}
\end{table}

The comparison of an optimal allocation with a worst allocation yields improvements that are about twice as large. Relative to a worst allocation, an optimal allocation improves test scores by 4.0 percent of a standard deviation on average (SE = 1.2 percent). The gains are 7.2 percent of a standard deviation for students with high baseline test scores, and 1.9 and 3.8 percent of a standard deviation for students with middle and low baseline test scores, respectively. If one considers only those students who are reassigned to a new teacher, the reallocation effect amounts 6.0 percent of a standard deviation on average (Panel B.II, SE = 1.9 percent).

One way to benchmark the magnitude of these AREs is to compare them with the effects of hypothetical policies aimed at improving teacher value-added measures (VAMs). As noted in the introduction, such policies are controversial, as is the evidence marshalled to support them. Here we offer no commentary on the advisability of actually adopting VAM-guided teacher personnel policies; nor do we offer an assessment of VAM studies. Rather we simply use these studies, and the policy thought experiments they motivate, to benchmark our ARE findings.

Teacher value-added is typically conceptualized as an invariant intercept-shifter, which uniformly raises or lowers the achievement of all students in a classroom. In this framework replacing a low value-added teacher with a high one will raise achievement for all students in a classroom. \citet{Rockoff_AER04} estimates that the standard deviation of the population distribution of teacher value-added (in a New Jersey school district) is around 0.10 test score standard deviations in both math and reading. Recent studies find somewhat higher estimates:~\citet{Chetty_et_al_AER2014a} estimates that the standard deviation of teacher value-added is 0.16 in math and 0.12 in reading; similarly, \citet{Rothstein2017} finds values of 0.19 in math and 0.12 in reading. 

Using a standard deviation of 0.15 we can consider the effect of a policy which removes the bottom $\tau \times 100$ percent of teachers, sorted by VAM, and replaces them with teachers at the $\tilde{\tau}^{th}$ quantile of the VAM distribution. Under normality the effect of such a policy on average student achievement is to increase test scores by
\begin{equation*}
    \left(1-\tau\right)\sigma\frac{\phi\left(\frac{q_{\tau}}{\sigma}\right)}{1-\Phi\left(\frac{q_{\tau}}{\sigma}\right)} + \sigma\Phi^{-1}\left(\tilde{\tau}\right)
\end{equation*}
standard deviations.
Setting $\tau = 0.05$ and $\tilde{\tau} = 0.75$ this expression gives an estimate of the policy effect of  0.021 (i.e., 2.1 percent of a test score standard deviation). This is comparable to the average effect on math achievement associated with moving from the status quo MET assignment to an optimal one. In practice correctly identifying, and removing from classrooms, the bottom five percent of teachers would be difficult to do. Replacing them with teachers in the top quartile of the VAM distribution perhaps even more so. Contextualized in this way the AREs we find are large.

An attractive feature of the policies we consider is that they are based on measurable student and teacher attributes, not noisily measured latent ones. At the same time we are mindful that most school districts would not find it costless to reallocate teachers freely across classrooms and schools.

Another way to calibrate the size of the effects we find is as follows. We find that when implementing an optimal teacher-to-classroom assignment, only about one half of students experience a change in teachers. We find that test scores increase by about 3.6 percent of a standard deviation for \emph{these} students. This is comparable to increasing the average VAM of these students' teachers from zero (i.e., the median) to the 0.6 quantile of the teacher VAM distribution. Again, increasing the VAM of half of all teachers by such an amount may be difficult to do in practice.\footnote{One may also use effect sizes of educational interventions in general as point of comparison. Based on a meta-study by~\citet{Kraft2020}, an effect size of 4 percent of a standard deviation in math qualifies as a ``medium effect,'' corresponding to the 40th percentile of effect sizes found across 750 educational interventions in pre-K--12 settings in the US (see Table 1 in~\citealp{Kraft2020}).} 

For English language arts (ELA) achievement we find smaller reallocation effects. Moving from the status quo to an optimal allocation is estimated to raise achievement by 0.8 percent of a test score standard deviation (SE = 0.6 percent). As with math, these gains are concentrated among students with high baseline scores who are assigned a new teacher. These students experience an average gain of 5 percent of a test score standard deviation (SE = 2.5 percent).

\begin{table}[h!tbp]
\centering
	\caption{Average reallocation gains in ELA. Gains expressed in  test score standard deviations.}
	\begin{small}
	\begin{tabular}{lccccccccc}
		\toprule
		\toprule
		& (1)        & (2)        & (3)        & (4)        &            & (5)        & (6)        & (7)        & (8) \\
		\midrule
		& \multicolumn{9}{c}{Panel A. Optimal versus status quo} \\
		& \multicolumn{4}{c}{A.I Full sample}               &            & \multicolumn{4}{c}{A.II Conditional on being reallocated} \\
		\cmidrule{2-5}\cmidrule{7-10}               & all students & high       & middle     & low        &            & all students & high       & middle     & low \\
		Gain       & 0.008      & 0.024      & 0.002      & 0.004      &            & 0.017      & 0.050      & 0.004      & 0.008 \\
		SE         & (0.006)    & (0.011)    & (0.007)    & (0.010)    &            & (0.012)    & (0.025)    & (0.013)    & (0.020) \\
		N          & 9,641      & 2,480      & 3,402      & 3,759      &            & 4,529      & 1,167      & 1,627      & 1,735 \\
		&            &            &            &            &            &            &            &            &  \\
		& \multicolumn{9}{c}{Panel B. Optimal versus worst allocation} \\
		\cmidrule{2-10}               & \multicolumn{4}{c}{B.I Full sample}               &            & \multicolumn{4}{c}{B.II Conditional on being reallocated} \\
		\cmidrule{2-5}\cmidrule{7-10}               & all students & high       & middle     & low        &            & all students & high       & middle     & low \\
		Gain       & 0.018      & 0.057      & 0.003      & 0.005      &            & 0.025      & 0.080      & 0.004      & 0.007 \\
		SE         & (0.010)    & (0.028)    & (0.011)    & (0.019)    &            & (0.015)    & (0.038)    & (0.017)    & (0.024) \\
		N          & 9,641      & 2,480      & 3,402      & 3,759      &            & 6,675      & 1,770      & 2,229      & 2,676 \\
		\bottomrule
		\bottomrule
	\end{tabular}%
	\end{small}
	\caption*{\footnotesize \textit{Note:} The table shows the average reallocation gains from implementing the optimal assignment instead of the random assignment (status quo) in Panel A, and the average reallocation gains from implementing the optimal assignment instead of the worst assignment in Panel B. The gains are expressed in test score standard deviations. The computations are based on a 3 $\times$ 3 model without teacher-by-peer interactions (see Table~\ref{tab:results_3-by-3_iv}, column 6). The results are presented separately for the full sample of students (columns 1--4), and for the sample of classrooms that would get a new teacher as a result of the reallocation (columns 5--8). The reassignments are carried out within school types and districts. Standard errors are in parentheses and computed using the Bayesian bootstrap with 1,000 replications (see Section~\ref{sec:inference}). High/middle/low: students in the top/middle/bottom tercile of the baseline test score distribution.}
\label{tab:reallocation-results_ela}
\end{table}%

In sum implementing an optimal assignment generates improvements in test score outcomes across the distribution of baseline achievement. Yet, while the gains are large for students with high baseline test scores \textendash{} up to 10 percent of a standard deviation for students with high baseline test scores in math \textendash{} for students with middle and low baseline test scores, the gains are smaller. Our results suggest that reallocations matter more in math, compared to ELA. This is in line with the value-added literature, which consistently reports higher value-added in math compared to ELA. 

Our results provide strong evidence of complementarity between student baseline test scores and teacher practices. This complementarity, if taken into account when assigning teachers to classrooms, appears to make a difference in students' performance across the distribution of baseline achievement. Teacher reassignments that maximize the average test score, however, will not close the achievement gap between students with low versus high baseline achievement levels. They actually widen this achievement gap. A different objective function, one that values equity as well as ``efficiency," may be preferred in practice.

It is also possible that assignments based on other student/teacher attributes might both raise average achievement and narrow achievement gaps. For example, if literacy in a student's native language helps develop literacy in English, then perhaps bilingual teachers could help to raise the achievement levels of English language learners. Many other examples might merit exploration.

\section{Sensitivity checks}
\label{sec:robustness}

\subsubsection*{Teacher and student categories} 

Working with a discrete categorization of baseline achievement and teacher FFT provides a simple, but also highly flexible, way of capturing non-linearities in educational production. We tested the sensitivity of our results to alternative categorizations of teachers and students. Specifically, we test (a) a coarser specification with two levels of teacher FFT (cutoff point at 2.5) and two levels of student baseline test scores and (b) a finer specification with four levels of teacher FFT (cutoff points at 2.25, 2.5, and 2.75) and four levels of student baseline test scores.  

In the coarser specification, we do not detect any significant match effects between teacher FFT and student baseline test scores (see Appendix Table~\ref{tab:results_2-by-2_iv}): the model with only two levels of teacher FFT does not capture the positive effect of the high-FFT teachers on students with middle or high baseline test scores. By contrast, the results of the finer specification, with four levels of FFT, are very similar to the results of our preferred specification~(see Appendix Table~\ref{tab:results_4-by-4_iv}). 

\subsubsection*{Measure of teaching practices} 

We also assess the sensitivity of our results to the measure of teaching practices that we use in the analysis. The MET data also contains an alternative measure, the CLASS (Classroom Assessment Scoring System), for a subset of the sample (6,320 observations in the math sample and 6,999 observations in the ELA sample). The CLASS is a teacher observation protocol that uses different domains and evaluation criteria than the FFT (see Section~\ref{sec:class} in the Data Appendix for details on the observation protocol and on how we process the data). The CLASS is also widely used in research on teacher quality (e.g.~\citealp{Carneiro2016}). We find that our results are similar when using the CLASS instead of the FFT (see Appendix Table~\ref{tab:results_3-by-3_class}).

\subsubsection*{Restrictions on the reassignment process}
In our preferred optimization procedure, we optimize the assignment of teachers to classrooms within types of schools (elementary or middle school) and school districts. We choose these restrictions because teachers might not be willing or able to teach in a different school type or a different school district. As a sensitivity check, we also calculated the results of both a more restrictive and a less restrictive optimization procedure. 

Our least restrictive allocation optimizes the assignment within school types, but allows for reassignments across districts. The magnitudes of the reallocation effects in this case are only slightly larger than those where reallocations are within districts (see Appendix Tables~\ref{tab:restrict-assignments_math} and~\ref{tab:restrict-assignments_ela}, columns 1--4). This finding may reflect the similarity of the distributions of baseline student achievement and teacher FFT in MET school districts. It would be interesting to repeat our analysis in a metro area consisting of an urban core district and multiple suburban ones. In such a setting is it seems plausible that moving teachers across school districts might raise average achievement.

Our most restrictive allocation reassigns teachers only within school-grade-subject cells (i.e., randomization blocks). This restriction is strong because each randomization block typically contains only two or three sections. Under this restriction, the reallocation gains are overall negligible (see Appendix Tables~\ref{tab:restrict-assignments_math} and~\ref{tab:restrict-assignments_ela}, columns 5--8).

\subsubsection*{Accounting for teacher-peer match effects in the assignment}

Our main results for the reallocation effects are based on a model without teacher-to-peer match effects (i.e., in a model which imposes the restriction that $\lambda = 0$), since these effects are very noisily estimated using our data. Using a model which does not impose this restriction generates reallocation effects about twice as large. These effects are also more noisily measured (see Appendix Tables~\ref{tab:teach-by-peer_math} and \ref{tab:teach-by-peer_ela}).

\section{Conclusion}

We provide an econometric framework that allows us to semiparametrically characterize complementarity between teaching practices and student baseline test scores. Our framework exploits the random assignment of teachers to classrooms available in the MET dataset, while formally dealing with non-compliance by both teachers and students.

Our results show that the potential gains associated with an outcome-maximizing assignment of teachers to classrooms are large. They are comparable to fairly large (hypothetical) interventions to raise teacher VAM. An attractive feature is that they are, at least in theory, resource neutral. No new teachers are required to implement the policies we consider. 

Our focus on the objective of maximizing the population average test score leads to an optimal assignment that increases the gap between less and more prepared students. We could also consider objective functions that do no focus on average test scores, but instead on test score gaps or proficiency levels. We could, for instance, choose an assignment which maximizes the number of students who reach a ``proficient" level on their end-of-year assessment. It is possible that this objective function would suggest a less assortative assignment: students with high baseline scores would likely reach the proficient level regardless of their teacher's FFT, while the incremental effect of having a high-FFT teacher on the probability of attaining proficiency may be larger for students with lower baseline test scores. Outcomes other than math and ELA achievement (e.g., socio-emotional skills) may also be of interest.

We consider this paper as a first pass that establishes the feasibility of recovering match effects from imperfect experimental data and shows that the resulting reallocation effects that depend upon these match effects and the supply of teachers are substantial.

\clearpage

\bibliography{met-1}

\clearpage

\appendix

\renewcommand{\thesection}{\Alph{section}}

\section{Figures and Tables}

\setcounter{figure}{0} \renewcommand{\thefigure}{A.\arabic{figure}} 
\setcounter{table}{0} \renewcommand{\thetable}{A.\arabic{table}}


\begin{figure}[h!tbp]
	\centering
	\caption{Distribution of teacher FFT and student baseline test scores}
	\includegraphics[width=0.7\linewidth]{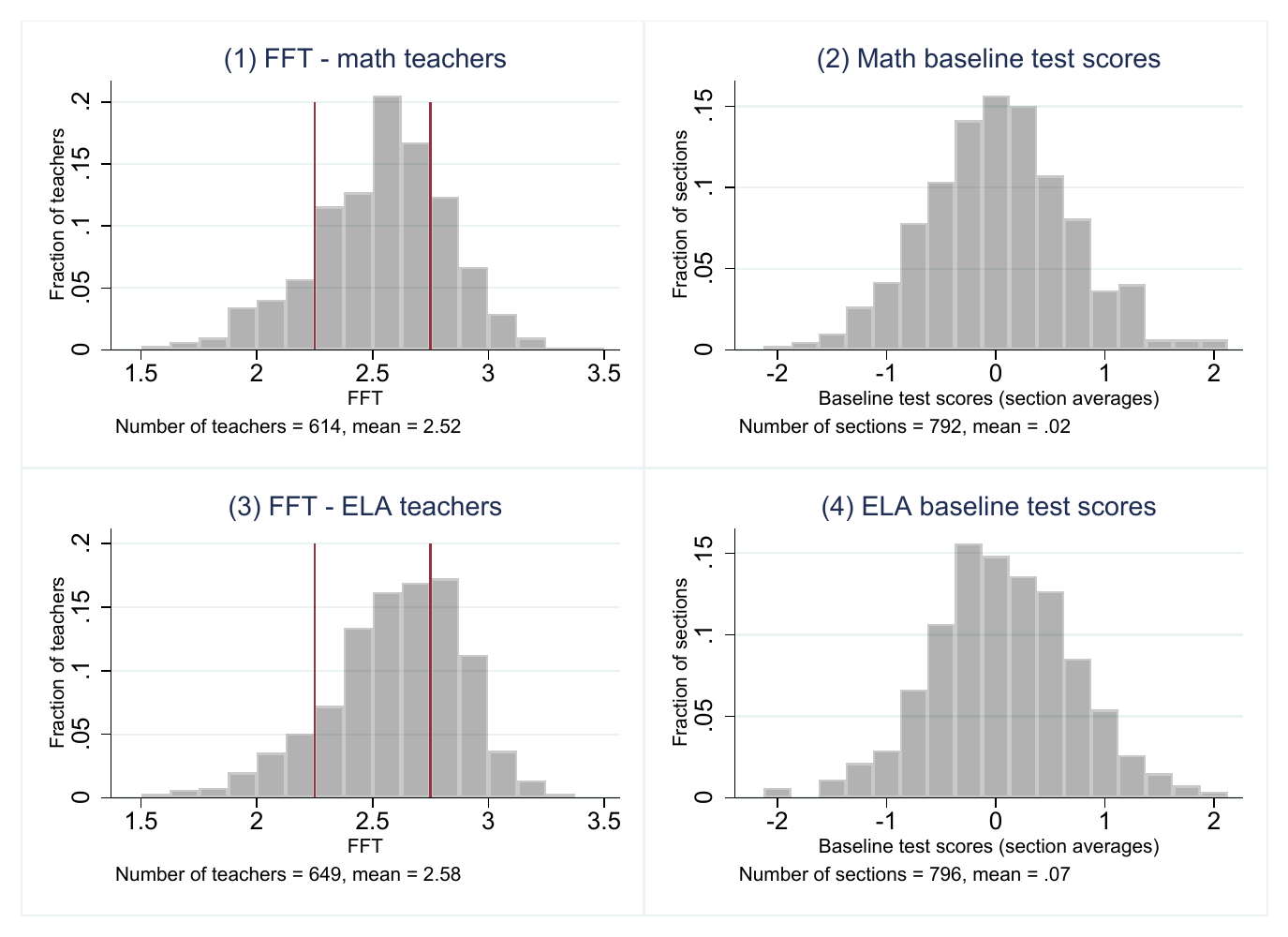}
	\label{fig:baseline-fft}
	\caption*{\footnotesize \textit{Note:} Distribution of teacher FFT (Panels 1 and 3) and student baseline test scores (Panels 2 and 4) in the estimation sample. Student baseline test scores are section averages.}
\end{figure}

\clearpage

\begin{figure}[h!tbp]
	\centering
	\caption{Assortativeness of the worst allocation in comparison with the status quo}
	\includegraphics[width=0.8\linewidth]{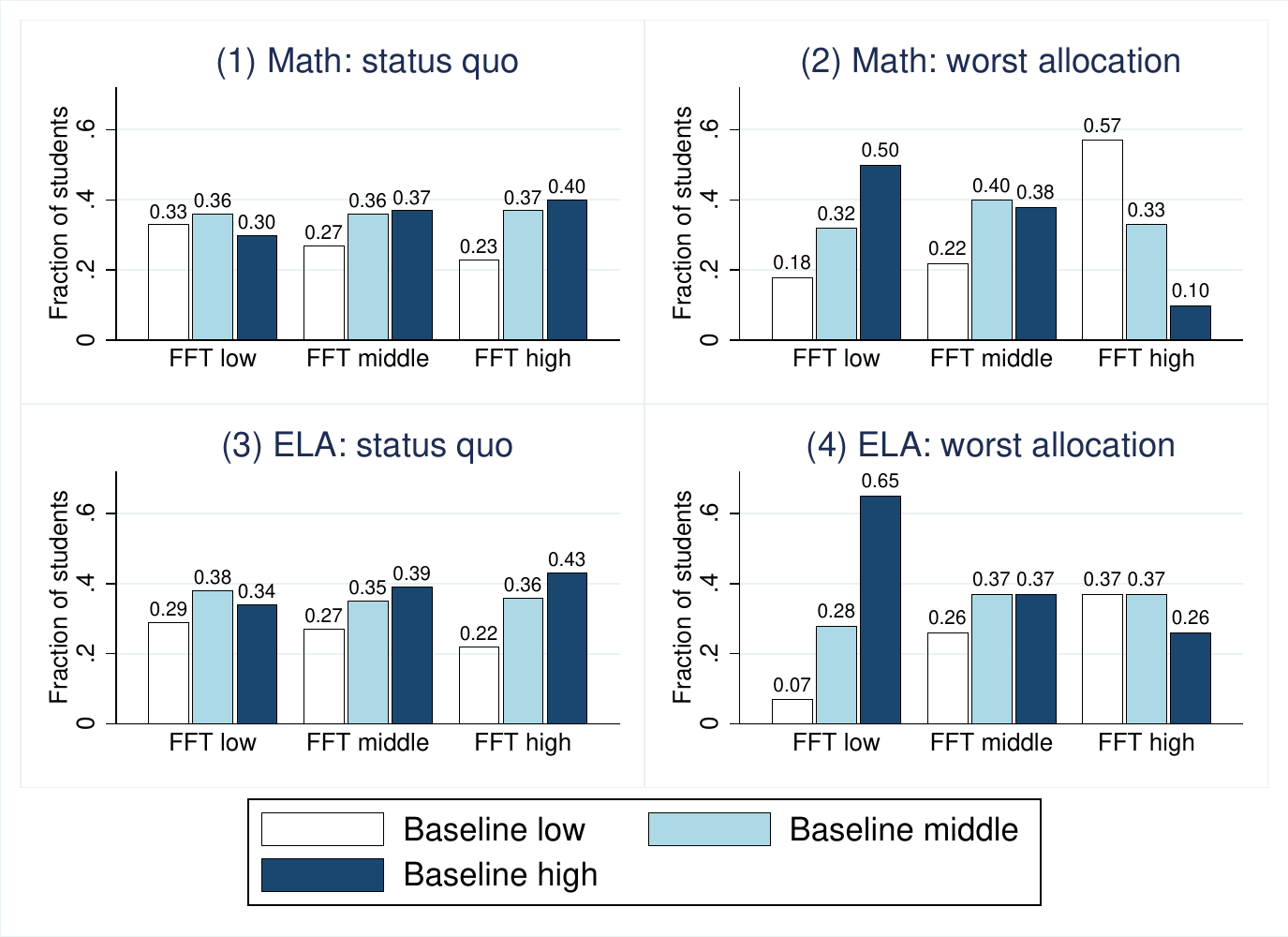}
	\label{fig:assortativeness-worst}
	\caption*{\footnotesize \textit{Note:} The figure compares the status quo allocation (Panels 1 and 3) with the worst allocation (Panels 2 and 4). The bars represent the fractions of students with low, middle, and high baseline test scores that a teacher with low, middle, and high FFT is assigned to on average under each of the allocations. For each teacher type, the fractions add up to 1. The optimization is carried out within school types (elementary or middle school) and districts.}
\end{figure}

\singlespacing
\begin{table}[h!tbp]
	\centering
	\caption{Summary statistics of the estimation samples}
\begin{small}
\begin{tabular}{rlrrrrr}
	\toprule
	\toprule
	&            & \multicolumn{1}{c}{(1)} & \multicolumn{1}{c}{(2)} &            & \multicolumn{1}{c}{(3)} & \multicolumn{1}{c}{(4)} \\
	\midrule
	&            & \multicolumn{2}{c}{Math sample} &            & \multicolumn{2}{c}{ELA sample} \\
	\cmidrule{3-4}\cmidrule{6-7}           &            & \multicolumn{1}{c}{Mean} & \multicolumn{1}{c}{SD} &            & \multicolumn{1}{c}{Mean} & \multicolumn{1}{c}{SD} \\
	\multicolumn{2}{l}{Student characteristics} &            &            &            &            &  \\
	& Age        & 10.32      & (1.56)     &            & 10.28      & (1.48) \\
	& Fourth grade & 26\%       & -          &            & 26\%       & - \\
	& Fifth grade & 35\%       & -          &            & 35\%       & - \\
	& Sixth grade & 16\%       & -          &            & 15\%       & - \\
	& Seventh grade & 10\%       & -          &            & 12\%       & - \\
	& Eighth grade & 12\%       & -          &            & 12\%       & - \\
	& Male       & 49\%       & -          &            & 49\%       & - \\
	& Gifted     & 6\%        & -          &            & 11\%       & - \\
	& Special education & 7\%        & -          &            & 7\%        & - \\
	& ELL        & 15\%       & -          &            & 13\%       & - \\
	& Free/reduced-price lunch & 58\%       & -          &            & 57\%       & - \\
	& White      & 27\%       & -          &            & 28\%       & - \\
	& Black      & 31\%       & -          &            & 30\%       & - \\
	& Hispanic   & 31\%       & -          &            & 33\%       & - \\
	& Asian      & 8\%        & -          &            & 7\%        & - \\
	& Other race & 3\%        & -          &            & 3\%        & - \\
	& Subject test score 2010 (baseline) & 0.11       & (0.89)     &            & 0.17       & (0.93) \\
	& Subject test score 2011 (outcome) & 0.15       & (0.90)     &            & 0.17       & (0.91) \\
	\multicolumn{2}{l}{Teacher characteristics} &            &            &            &            &  \\
	& Male       & 14\%       & -          &            & 11\%       & - \\
	& White      & 65\%       & -          &            & 64\%       & - \\
	& Black      & 28\%       & -          &            & 29\%       & - \\
	& Hispanic   & 6\%        & -          &            & 6\%        & - \\
	& Other race & 2\%        & -          &            & 1\%        & - \\
	& Years in district & 7.39       & (6.73)     &            & 7.36       & (6.33) \\
	& Master's or higher & 40\%       & -          &            & 37\%       & - \\
	& FFT        & 2.53       & (0.30)     &            & 2.59       & (0.30) \\
	\multicolumn{2}{l}{Classroom characterisics} &            &            &            &            &  \\
	& Class size & 25.54      & (5.48)     &            & 25.60      & (5.44) \\
	\multicolumn{2}{l}{Sample size} &            &            &            &            &  \\
	& Students   & \multicolumn{2}{c}{8,534} &            & \multicolumn{2}{c}{9,641} \\
	& Teachers   & \multicolumn{2}{c}{614} &            & \multicolumn{2}{c}{649} \\
	& Schools    & \multicolumn{2}{c}{153} &            & \multicolumn{2}{c}{160} \\
	\bottomrule
	\bottomrule
\end{tabular}%
\end{small}
	\caption*{\footnotesize \textit{Note:} Summary statistics of the estimation samples. Standard deviations are in parentheses. ELL: English language learner. FFT: Framework for teaching. For details on the sample construction, see Appendix~\ref{sec:data_appendix}.}
	\label{tab:descriptives}
\end{table}

\begin{table}[H]
	\centering
	\caption{Balancing tests. Dependent variable: FFT of the assigned teacher}
\begin{small}
\begin{tabular}{lrrrrr}
	\toprule
	\toprule
	& (1)        & (2)        &            & (3)        & (4) \\
	\midrule
	& \multicolumn{2}{c}{Math} &            & \multicolumn{2}{c}{ELA} \\
	\cmidrule{2-3}\cmidrule{5-6}           & \multicolumn{1}{c}{coeff}      & \multicolumn{1}{c}{SE}         &            & \multicolumn{1}{c}{coeff}      & \multicolumn{1}{c}{SE}  \\
	Baseline test score & 0.004      & (0.005)    &            & 0.000      & (0.004) \\
	Age        & -0.003     & (0.006)    &            & -0.006     & (0.005) \\
	Male       & 0.001      & (0.005)    &            & -0.002     & (0.003) \\
	Gifted     & -0.015     & (0.015)    &            & 0.010      & (0.016) \\
	Special education & 0.013      & (0.012)    &            & -0.002     & (0.009) \\
	ELL        & -0.004     & (0.011)    &            & -0.014     & (0.011) \\
	Black      & -0.004     & (0.008)    &            & -0.010     & (0.007) \\
	Hispanic   & -0.013     & (0.009)    &            & 0.002      & (0.006) \\
	Asian      & 0.009      & (0.014)    &            & 0.016      & (0.011) \\
	Other race & 0.014      & (0.012)    &            & 0.010      & (0.010) \\
	Free/reduced-price lunch & 0.005      & (0.010)    &            & -0.002     & (0.006) \\
	\midrule
	\multicolumn{1}{p{11.07em}}{F-test for joint significance (p-value)} & \multicolumn{2}{c}{0.465} &            & \multicolumn{2}{c}{0.241} \\
	$R^2$       & \multicolumn{2}{c}{0.598} &            & \multicolumn{2}{c}{0.635} \\
	N          & \multicolumn{2}{c}{8,518} &            & \multicolumn{2}{c}{9,630} \\
	\bottomrule
	\bottomrule
\end{tabular}%
\end{small}
	\caption*{\footnotesize \textit{Note:} The table presents results of OLS regressions of the FFT of the assigned teacher on individual student characteristics. Free/reduced-price lunch eligibility is coded as 0 for students who do not have information on this variable in the dataset. All regressions control for randomization block fixed effects. Analytic standard errors, clustered by randomization block, are in parentheses. ELL: English language learner. FFT: Framework for teaching.}
	\label{tab:balancing}
\end{table}

\begin{table}[H]
	\centering
	\caption{Tests of Assumption~\ref{ass:key_ass_1}. Dependent variables: Differences between assigned and realized teacher characteristics}
\begin{small}
\begin{tabular}{lcccccc}
	\toprule
	\toprule
	& (1)        & (2)        & (3)        & (4)        & (5)        & (6) \\
	\midrule
	&            &            &            &            &            &  \\
	& \multicolumn{6}{c}{Panel I. Math sample} \\
	\cmidrule{2-7}           & Male       & \multicolumn{3}{c}{Teacher's race}   & Experience & Master's \\
	&            & white      & black      & hispanic   & (years)    & degree \\
	\midrule
    	FFT, assigned teacher & 0.009      & -0.039     & -0.061     & 0.059      & -2.727     & -0.018 \\
	& (0.060)    & (0.107)    & (0.068)    & (0.059)    & (2.236)    & (0.099) \\
	Baseline test score & 0.003      & -0.003     & 0.005      & -0.002     & 0.071      & -0.006 \\
	& (0.005)    & (0.005)    & (0.004)    & (0.002)    & (0.082)    & (0.006) \\
	Avg. peer baseline test score, & 0.027      & 0.037      & 0.007      & -0.037    &  0.738      & -0.052 \\
	assigned peers & (0.076)    & (0.058)    & (0.034)    & (0.041)    & (0.931)    & (0.089) \\
	\midrule
	\multicolumn{1}{p{12em}}{F-test for joint significance (p-value)} & 0.956      & 0.799      & 0.537      & 0.731      & 0.635      & 0.849 \\
	$R^2$  & 0.149      & 0.176      & 0.108      & 0.039      & 0.163      & 0.154 \\
	N          & 8,163      & 8,098      & 7,689      & 7,689      & 5,585      & 6,083 \\
	\midrule
	&            &            &            &            &            &  \\
	& \multicolumn{6}{c}{Panel II. ELA sample} \\
	\cmidrule{2-7}           & Male       & \multicolumn{3}{c}{Teacher's race}   & Experience & Master's \\
	&            & white      & black      & hispanic   & (years)    & degree \\
	\midrule
    	FFT, assigned teacher & 0.008      & 0.036      & -0.037     & 0.031      & 0.500      & -0.003 \\
	& (0.067)    & (0.085)    & (0.071)    & (0.028)    & (0.799)    & (0.060) \\
	Baseline test score & 0.003      & -0.000     & 0.002      & -0.002     & -0.008     & 0.001 \\
	& (0.002)    & (0.004)    & (0.003)    & (0.002)    & (0.042)    & (0.004) \\
	Avg. peer baseline test score, & -0.008     & 0.102**    & -0.041     & -0.056*    & 0.220      & 0.049 \\
	assigned peers & (0.027)    & (0.046)    & (0.031)    & (0.029)    & (0.436)    & (0.049) \\
	\midrule
	\multicolumn{1}{p{12em}}{F-test for joint significance (p-value)} & 0.300      & 0.116      & 0.414      & 0.063      & 0.837      & 0.779 \\
	$R^2$  & 0.090      & 0.130      & 0.053      & 0.050      & 0.142      & 0.111 \\
	N          & 9,183      & 9,183      & 9,183      & 9,183      & 6,791      & 6,963 \\
	\bottomrule
	\bottomrule
\end{tabular}%
\end{small}
	\caption*{\footnotesize \textit{Note:} The table presents results of OLS regressions of differences between assigned and realized teacher characteristics on individual baseline test scores, the FFT of the assigned teacher, and the average baseline test scores of the assigned peers. Each column represents a regression with a different dependent variable. All regressions control for randomization block fixed effects. Analytic standard errors, clustered by randomization block, are in parentheses. Teachers with ethnicity "white"  and ethnicity "other" are pooled into one category. FFT: Framework for teaching. \\ **significant at the 5\%-level, *significant at the 10\%-level.}
	\label{tab:assumption-1}
\end{table}

\begin{landscape}
	\begin{table}[H]
		\centering
		\caption{Tests of Assumption~\ref{ass:key_ass_2} in the math sample. Dependent variables: differences between assigned and realized peer characteristics}
\begin{small}
\begin{tabular}{lcccccccccc}
\toprule
\toprule
        & \multicolumn{1}{c}{(1)} & \multicolumn{1}{c}{(2)} & \multicolumn{1}{c}{(3)} & \multicolumn{1}{c}{(4)} & \multicolumn{1}{c}{(5)} & \multicolumn{1}{c}{(6)} & \multicolumn{1}{c}{(7)} & \multicolumn{1}{c}{(8)} & \multicolumn{1}{c}{(9)} & \multicolumn{1}{c}{(10)} \\
\midrule
\multicolumn{1}{p{14.43em}}{Difference between assigned } & \multicolumn{1}{c}{Age} & \multicolumn{1}{c}{Male} & \multicolumn{1}{c}{Gifted} & \multicolumn{1}{c}{Special} & \multicolumn{1}{c}{ELL} & \multicolumn{1}{c}{FRL} & \multicolumn{4}{c}{Race/ethnicity} \\
\cmidrule{8-11}and realized peer characteristics &         &         &         & \multicolumn{1}{c}{education} &         &         & \multicolumn{1}{c}{White} & \multicolumn{1}{c}{Black} & \multicolumn{1}{c}{Hispanic} & \multicolumn{1}{c}{Asian} \\
\midrule
FFT, assigned teacher & 0.008   & 0.015   & 0.006   & -0.017  & 0.006   & -0.005  & 0.004   & -0.010  & 0.006   & -0.001 \\
        & (0.019) & (0.014) & (0.009) & (0.011) & (0.012) & (0.016) & (0.011) & (0.011) & (0.013) & (0.007) \\
Baseline test score & -0.006  & 0.000   & 0.004** & -0.001  & 0.000   & 0.000   & 0.001   & -0.002  & 0.000   & 0.001* \\
        & (0.003) & (0.001) & (0.002) & (0.001) & (0.001) & (0.002) & (0.001) & (0.002) & (0.001) & (0.001) \\
Avg. peer baseline test score,  & -0.030* & -0.006  & -0.013  & 0.009   & 0.033** & -0.001  & -0.012  & 0.012   & -0.007  & 0.008 \\
assigned peers & (0.017) & (0.013) & (0.009) & (0.008) & (0.011) & (0.014) & (0.010) & (0.011) & (0.011) & (0.005) \\
\midrule
$R^2$ & 0.124   & 0.264   & 0.222   & 0.240   & 0.200   & 0.241   & 0.207   & 0.232   & 0.247   & 0.185 \\
N       & 8,525   & 8,534   & 8,534   & 8,527   & 8,534   & 5,976   & 8,534   & 8,534   & 8,534   & 8,534 \\
\bottomrule
\bottomrule
\end{tabular}%
\end{small}

		\caption*{\footnotesize \textit{Note:} The table presents results of OLS regressions of differences between assigned and realized peer characteristics on the FFT of the assigned teacher, individual baseline test scores, and the average baseline test scores of the assigned peers. Each column represents a regression with a different dependent variable. All regressions control for randomization block fixed effects. Analytic standard errors, clustered by randomization block, are in parentheses. ELL: English language learner, FRL: Free/reduced-price lunch eligible, FFT: Framework for teaching. \\ ** significant at the 5\%-level * significant at the 10\%-level.}
		\label{tab:assumption-2_math}
	\end{table}
\end{landscape}

\begin{landscape}
	\begin{table}[H]
		\centering
		\caption{Tests of Assumption~\ref{ass:key_ass_2} in the ELA sample. Dependent variables: differences between assigned and realized peer characteristics}
\begin{small}
\begin{tabular}{lcccccccccc}
\toprule
\toprule
        & \multicolumn{1}{c}{(1)} & \multicolumn{1}{c}{(2)} & \multicolumn{1}{c}{(3)} & \multicolumn{1}{c}{(4)} & \multicolumn{1}{c}{(5)} & \multicolumn{1}{c}{(6)} & \multicolumn{1}{c}{(7)} & \multicolumn{1}{c}{(8)} & \multicolumn{1}{c}{(9)} & \multicolumn{1}{c}{(10)} \\
\midrule
\multicolumn{1}{p{14.43em}}{Difference between assigned } & \multicolumn{1}{c}{Age} & \multicolumn{1}{c}{Male} & \multicolumn{1}{c}{Gifted} & \multicolumn{1}{c}{Special} & \multicolumn{1}{c}{ELL} & \multicolumn{1}{c}{FRL} & \multicolumn{4}{c}{Race/ethnicity} \\
\cmidrule{8-11}and realized peer characteristics &         &         &         & \multicolumn{1}{c}{education} &         &         & \multicolumn{1}{c}{White} & \multicolumn{1}{c}{Black} & \multicolumn{1}{c}{Hispanic} & \multicolumn{1}{c}{Asian} \\
\midrule
FFT,  assigned teacher & 0.000   & 0.008   & 0.008   & -0.012  & 0.022   & -0.003  & -0.008  & -0.001  & 0.012   & -0.002 \\
        & (0.020) & (0.014) & (0.009) & (0.011) & (0.017) & (0.014) & (0.012) & (0.013) & (0.014) & (0.009) \\
Baseline test score & -0.005** & -0.001  & 0.002   & 0.000   & -0.001  & -0.002  & 0.001   & 0.000   & -0.001  & 0.001 \\
        & (0.003) & (0.001) & (0.001) & (0.001) & (0.001) & (0.001) & (0.001) & (0.001) & (0.001) & (0.000) \\
Avg. peer baseline test score,  & -0.009  & 0.000   & -0.029*** & 0.013   & 0.028** & 0.007   & -0.002  & 0.001   & 0.001   & 0.000 \\
assigned peers & (0.013) & (0.011) & (0.007) & (0.009) & (0.009) & (0.011) & (0.008) & (0.008) & (0.010) & (0.006) \\
\midrule
$R^2$ & 0.098   & 0.241   & 0.147   & 0.208   & 0.178   & 0.238   & 0.203   & 0.254   & 0.218   & 0.215 \\
N       & 9,635   & 9,641   & 9,641   & 9,636   & 9,641   & 7,270   & 9,641   & 9,641   & 9,641   & 9,641 \\
\bottomrule
\bottomrule
\end{tabular}%
\end{small}

		\caption*{\footnotesize \textit{Note:} The table presents results of OLS regressions of differences between assigned and realized peer characteristics on the FFT of the assigned teacher, individual baseline test scores, and the average baseline test scores of the assigned peers. Each column represents a regression with a different dependent variable. All regressions control for randomization block fixed effects. Analytic standard errors, clustered by randomization block, are in parentheses. ELL: English language learner, FRL: Free/reduced-price lunch eligible, FFT: Framework for teaching. \\ ** significant at the 5\%-level * significant at the 10\%-level.}
		\label{tab:assumption-2_ela}
	\end{table}
\end{landscape}

\begin{table}[H]
\centering
\caption{Tests of restriction~(\ref{eq: key_ass_2 obs}). Dependent variables: average peer baseline test scores of realized peers}
\begin{small}
\begin{tabular}{lccc}
\toprule
\toprule
       & (1)    &        & (2) \\
\midrule
       & \multicolumn{3}{c}{Average peer baseline test score,} \\
       & \multicolumn{3}{c}{realized peers} \\
\cmidrule{2-4}       & Math sample &        & ELA sample \\
\cmidrule{1-2}\cmidrule{4-4} 
FFT, assigned teacher & -0.015 &        & 0.036 \\
       & (0.029) &        & (0.029) \\
Baseline test score & 0.031*** &        & 0.023*** \\
       & (0.008) &        & (0.005) \\
Avg. peer baseline test score,  & 0.744  &        & 0.769 \\
assigned peers & (0.034) &        & (0.030) \\
\midrule
$R^2$ & 0.881  &        & 0.904 \\
N      & 8,534  &        & 9,641 \\
\bottomrule
\bottomrule
\end{tabular}%
\end{small}
	\caption*{\footnotesize \textit{Note:} The table presents results of OLS regressions of the average peer baseline test scores of the realized peers on individual baseline test scores, FFT of the assigned teacher, and the average peer baseline test scores of the assigned peers in the math and ELA samples. All regressions control for randomization block fixed effects. Analytic standard errors, clustered by randomization block, are in parentheses. FFT: Framework for teaching. \\ *** significant at the 1\%-level.}
	\label{tab:assumption-3}
\end{table}

\clearpage

\begin{table}[htbp]
  \centering
  \caption{Tests for instrument relevance}
  \begin{small}
    \begin{tabular}{p{16.715em}rrrr}
    \hline
    \hline
    \multicolumn{1}{r}{} & \multicolumn{1}{c}{(1)} & \multicolumn{1}{c}{(2)} &         & \multicolumn{1}{c}{(3)} \bigstrut[t]\\
    \multicolumn{5}{l}{Panel A. Math} \bigstrut[b]\\
    \hline
    \multicolumn{1}{r}{} & \multicolumn{4}{c}{First-stage F-statistics} \bigstrut[t]\\
    \multicolumn{1}{r}{} & \multicolumn{2}{c}{unconditional } &         & \multicolumn{1}{c}{SW} \bigstrut[b]\\
\cline{2-3}\cline{5-5}    Variable & \multicolumn{1}{p{4.145em}}{F(10,238)} & \multicolumn{1}{p{3.855em}}{p-value} &         & \multicolumn{1}{p{5.145em}}{ F(1, 238)} \bigstrut[t]\\
    \multicolumn{1}{r}{} &         &         &         &  \\
    FFT middle & 17.57   & (0.000) &         & 60.17 \\
    FFT high & 11.62   & (0.000) &         & 57.95 \\
    FFT middle $\times$ baseline middle & 82.49   & (0.000) &         & 498.96 \\
    FFT high $\times$ baseline middle & 33.30   & (0.000) &         & 387.27 \\
    FFT middle $\times$ baseline high & 39.59   & (0.000) &         & 322.43 \\
    FFT high $\times$ baseline high & 36.92   & (0.000) &         & 378.59 \\
    FFT middle $\times$ fraction peers middle & 17.47   & (0.000) &         & 113.27 \\
    FFT high $\times$ fraction peers middle & 12.85   & (0.000) &         & 102.70 \\
    FFT middle $\times$ fraction peers high & 8.07    & (0.000) &         & 182.12 \\
    FFT middle $\times$ fraction peers high & 12.01   & (0.000) &         & 137.55 \\
    \multicolumn{1}{r}{} &         &         &         &  \\
    \multicolumn{5}{l}{Panel B. ELA} \bigstrut[b]\\
    \hline
    \multicolumn{1}{r}{} & \multicolumn{4}{c}{First-stage F-statistics} \bigstrut[t]\\
    \multicolumn{1}{r}{} & \multicolumn{2}{c}{unconditional } &         & \multicolumn{1}{c}{SW} \bigstrut[b]\\
\cline{2-3}\cline{5-5}    Variable & \multicolumn{1}{p{4.145em}}{F(10,238)} & \multicolumn{1}{p{3.855em}}{p-value} &         & \multicolumn{1}{p{5.145em}}{ F(1, 238)} \bigstrut[t]\\
    \multicolumn{1}{r}{} &         &         &         &  \\
    FFT middle & 26.21   & (0.000) &         & 75.61 \\
    FFT high & 22.34   & (0.000) &         & 102.87 \\
    FFT middle $\times$ baseline middle & 146.86  & (0.000) &         & 836.59 \\
    FFT high $\times$ baseline middle & 121.72  & (0.000) &         & 1500.06 \\
    FFT middle $\times$ baseline high & 128.34  & (0.000) &         & 1226.48 \\
    FFT high $\times$ baseline high & 88.44   & (0.000) &         & 1724.38 \\
    FFT middle $\times$ fraction peers middle & 25.06   & (0.000) &         & 164.19 \\
    FFT high $\times$ fraction peers middle & 21.57   & (0.000) &         & 213.78 \\
    FFT middle $\times$ fraction peers high & 13.9    & (0.000) &         & 362.49 \\
    FFT middle $\times$ fraction peers high & 13.21   & (0.000) &         & 477.11 \bigstrut[b]\\
    \hline
    \hline
    \end{tabular}%
    \end{small}
  \label{tab:firststage}%
  \caption*{\footnotesize \textit{Note:} The table presents weak instrument F-tests following~\citet{Sanderson2016}. Columns 1 and 2 present the unconditional F-statistics from first-stage regressions of equation~(\ref{eq: plm_endogenous_version}). Column 3 presents the Sanderson-Windmeijer conditional F-statistics for each of the endogenous variables. The F-statistics in column 3 need to be compared to the critical values of the Stock-Yogo weak identification F-test~\citep{Stock2005}. All of the F-statistics exceed the Stock-Yogo critical values for a 5\% maximal IV relative bias (critical value is 20.74).}
\end{table}%

\clearpage

\begin{table}[H]
	\centering
	\caption{OLS regression results of the linear model. Dependent variables: student test score outcomes}
\begin{small}
\begin{tabular}{lrcccccc}
	\toprule
	\toprule
	&            & (1)        & (2)        & (3)        & (4)        & (5)        & (6) \\
	&            & \multicolumn{2}{c}{A. Only teacher} & \multicolumn{2}{c}{B. Full model} & \multicolumn{2}{c}{C. Without} \\
	&            & \multicolumn{2}{c}{effects} &            &            & \multicolumn{2}{c}{teacher $\times$ peer} \\
	&            & \multicolumn{2}{c}{}    &            &            & \multicolumn{2}{c}{interactions} \\
	&            & Math       & ELA        & Math       & ELA        & Math       & ELA \\
	\hline
	$\delta$ & \multicolumn{1}{l}{FFT } & 0.088**    & 0.118**    & 0.081**    & 0.116**    & 0.080**    & 0.116** \\
	&            & (0.040)    & (0.045)    & (0.040)    & (0.045)    & (0.040)    & (0.044) \\
	$\eta$ & \multicolumn{1}{l}{FFT $\times$ baseline} &            &            & 0.060*     & 0.088**    & 0.059**    & 0.078** \\
	&            &            &            & (0.031)    & (0.032)    & (0.028)    & (0.030) \\
	$\lambda$ & \multicolumn{1}{l}{FFT $\times$ avg. peer baseline} &            &            & -0.006     & -0.066     &            &  \\
	&            &            &            & (0.066)    & (0.081)    &            &  \\
	$\beta$ & \multicolumn{1}{l}{Baseline } & 0.749***   & 0.690***   & 0.595***   & 0.460***   & 0.598***   & 0.487*** \\
	&            & (0.011)    & (0.011)    & (0.079)    & (0.081)    & (0.073)    & (0.078) \\
	\midrule
	$R^2$  &            & 0.700      & 0.633      & 0.701      & 0.633      & 0.701      & 0.633 \\
	N          &            & 8,534      & 9,641      & 8,534      & 9,641      & 8,534      & 9,641 \\
	\bottomrule
	\bottomrule
\end{tabular}%
\end{small}
	\caption*{\footnotesize \textit{Note:} The dependent variables are subject-specific test score outcomes in math and ELA. The specifications include linear terms for FFT, individual and peer baseline test scores. All regressions control for the $h(x,\overline{x})$ function (see Section~\ref{sec:model}) and for randomization block fixed effects. Analytic standard errors, clustered by randomization block, are in parentheses. FFT: Framework for teaching. \\ *** significant at the 1\%-level ** significant at the 5\%-level * significant at the 10\%-level.}
	\label{tab:results_lin_ols}
\end{table}


\begin{table}[H]
	\centering
	\caption{OLS regression results of the 3 $\times$ 3 model. Dependent variables: student test score outcomes}
\begin{small}
\begin{tabular}{lrcccccc}
	\toprule
	\toprule
	&            & (1)        & (2)        & (3)        & (4)       & (5)       & (6) \\
	&            & \multicolumn{2}{c}{A. Only teacher} & \multicolumn{2}{c}{B. Full model} & \multicolumn{2}{c}{C. Without} \\
	&            & \multicolumn{2}{c}{effects} &            &            & \multicolumn{2}{c}{teacher $\times$ peer} \\
	&            & \multicolumn{2}{c}{}    &            &            & \multicolumn{2}{c}{interactions} \\
	&            & Math       & ELA        & Math       & ELA        & Math       & ELA \\
	\midrule
	$\delta$   & \multicolumn{1}{l}{FFT middle} & 0.062*     & 0.028      & 0.032      & -0.058     & 0.032      & -0.127** \\
	&            & (0.032)    & (0.034)    & (0.097)    & (0.091)    & (0.038)    & (0.045) \\
	& \multicolumn{1}{l}{FFT high} & 0.066      & 0.048      & -0.060     & -0.043     & -0.016     & -0.141** \\
	&            & (0.041)    & (0.043)    & (0.113)    & (0.125)    & (0.052)    & (0.054) \\
	$\eta$        & \multicolumn{1}{l}{FFT middle } &            &            &            &            &            &  \\
	& \multicolumn{1}{l}{\hspace{0.1cm} $\times$  baseline middle} &            &            & 0.030      & 0.108**    & 0.033      & 0.102** \\
	&            &            &            & (0.047)    & (0.054)    & (0.045)    & (0.051) \\
	& \multicolumn{1}{l}{\hspace{0.1cm} $\times$ baseline high} &            &            & 0.039      & 0.136**    & 0.053      & 0.137** \\
	&            &            &            & (0.060)    & (0.067)    & (0.055)    & (0.062) \\
	& \multicolumn{1}{l}{FFT high} &            &            &            &            &            &  \\
	& \multicolumn{1}{l}{\hspace{0.1cm} $\times$ baseline middle} &            &            & 0.100*     & 0.113*     & 0.098*     & 0.094 \\
	&            &            &            & (0.059)    & (0.060)    & (0.059)    & (0.059) \\
	& \multicolumn{1}{l}{\hspace{0.1cm} $\times$ baseline high} &            &            & 0.120*     & 0.225**    & 0.104      & 0.193** \\
	&            &            &            & (0.068)    & (0.075)    & (0.067)    & (0.072) \\
	$\lambda$     & \multicolumn{1}{l}{FFT middle} &            &            &            &            &            &  \\
	& \multicolumn{1}{l}{\hspace{0.1cm} $\times$ fraction peers middle} &            &            & -0.024     & -0.204     &            &  \\
	&            &            &            & (0.198)    & (0.187)    &            &  \\
	& \multicolumn{1}{l}{\hspace{0.1cm} $\times$ fraction peers high} &            &            & -0.045     & -0.126     &            &  \\
	&            &            &            & (0.187)    & (0.154)    &            &  \\
	& \multicolumn{1}{l}{FFT high} &            &            &            &            &            &  \\
	& \multicolumn{1}{l}{\hspace{0.1cm} $\times$ fraction peers middle} &            &            & 0.155      & -0.187     &            &  \\
	&            &            &            & (0.224)    & (0.232)    &            &  \\
	& \multicolumn{1}{l}{\hspace{0.1cm} $\times$ fraction peers high} &            &            & 0.460      & -0.355     &            &  \\
	&            &            &            & (0.409)    & (0.343)    &            &  \\
	&            &            &            &            &            &            &  \\
	$\beta$       & \multicolumn{1}{l}{Baseline middle} & 0.911***   & 0.823***   & 0.782***   & 0.713***   & 0.783***   & 0.723*** \\
	&            & (0.072)    & (0.075)    & (0.076)    & (0.093)    & (0.075)    & (0.095) \\
	& \multicolumn{1}{l}{Baseline high} & 1.678***   & 1.636***   & 1.551***   & 1.443***   & 1.544***   & 1.454*** \\
	&            & (0.105)    & (0.101)    & (0.111)    & (0.113)    & (0.106)    & (0.116) \\
	\midrule
	$R^2$        &            & 0.618      & 0.557      & 0.570      & 0.523      & 0.570      & 0.523 \\
	N          &            & 8,534      & 9,641      & 8,534      & 9,641      & 8,534      & 9,641 \\
	\bottomrule
	\bottomrule
\end{tabular}%
\end{small}
	\caption*{\footnotesize \textit{Note:} The dependent variables are subject-specific test score outcomes in math and ELA. All regressions control for the $h(x, \overline{x})$ function (see Section~\ref{sec:model}) and for randomization block fixed effects. Analytic standard errors, clustered by randomization block, are in parentheses. FFT: Framework for teaching. \\ *** significant at the 1\%-level ** significant at the 5\%-level * significant at the 10\%-level.}
	\label{tab:results_3-by-3_ols}
\end{table}



\begin{table}[H]
	\centering
	\caption{IV regression results of the 2 $\times$ 2 model. Dependent variables: student test score outcomes}
\begin{small}
\begin{tabular}{lrcccccc}
	\toprule
	\toprule
	&            & (1)        & (2)        & (3)        & (4)        & (5)        & (6) \\
	&            & \multicolumn{2}{c}{A. Only teacher} & \multicolumn{2}{c}{B. Full model} & \multicolumn{2}{c}{C. Without } \\
	&            &  \multicolumn{2}{c}{effects}                  &            &            & \multicolumn{2}{c}{teacher $\times$ peer} \\
	&            &            &            &            &            & \multicolumn{2}{c}{interactions} \\
	&            & Math       & ELA        & Math       & ELA        & Math       & ELA \\
	\hline
	$\delta$   & \multicolumn{1}{l}{FFT high} & 0.072*     & -0.060     & 0.088      & 0.170      & 0.056      & -0.027 \\
	&            & (0.038)    & (0.043)    & (0.087)    & (0.107)    & (0.047)    & (0.046) \\
	$\eta$     & \multicolumn{1}{l}{\hspace{0.1cm}$\times$  baseline high} &            &            & 0.039      & -0.014     & 0.030      & -0.062 \\
	&  &            &            & (0.053)    & (0.055)    & (0.052)    & (0.055) \\
	$\lambda$  & \multicolumn{1}{l}{\hspace{0.1cm}$\times$  fraction peers high} &            &            & -0.068     & -0.422*    &            &  \\
	& &            &            & (0.149)    & (0.229)    &            &  \\
	$\beta$    & \multicolumn{1}{l}{Baseline high} & 0.935***   & 0.927***   & 0.920***   & 0.938***   & 0.924***   & 0.956*** \\
	&            & (0.058)    & (0.054)    & (0.058)    & (0.060)    & (0.057)    & (0.060) \\
	\midrule
	$R^2$      &            & 0.545      & 0.485      & 0.545      & 0.481      & 0.545      & 0.485 \\
	N          &            & 8,534      & 9,641      & 8,534      & 9,641      & 8,534      & 9,641 \\
	\bottomrule
	\bottomrule
\end{tabular}%
\end{small}
	\caption*{\footnotesize \textit{Note:} The dependent variables are subject-specific test score outcomes in math and ELA. The instrumental variables are based on assigned teacher FFT (Panels A--C) and assigned peer baseline test scores (Panel B).  All regressions control for the $h(x,\overline{x})$ function (see Section~\ref{sec:model}) and for randomization block fixed effects. Analytic standard errors, clustered by randomization block, are in parentheses. FFT: Framework for teaching. \\ *** significant at the 1\%-level ** significant at the 5\%-level * significant at the 10\%-level.}
	\label{tab:results_2-by-2_iv}
\end{table}

\begin{table}[H]
	\centering
	\caption{IV regression results of the 4 $\times$ 4 model. Dependent variables: student test score outcomes}
\begin{small}
\begin{tabular}{lrcccccc}
	\toprule
	\toprule
	&            & (1)        & (2)        & (3)        & (4)        & (5)        & (6) \\
	&            & \multicolumn{2}{c}{A. Only teacher } & \multicolumn{2}{c}{B. Full model} & \multicolumn{2}{c}{C. Without } \\
	&            & \multicolumn{2}{c}{effects} &            &            & \multicolumn{2}{c}{teacher $\times$ peer} \\
	&            &            &            &            &            & \multicolumn{2}{c}{interactions} \\
		&            & Math       & ELA        & Math       & ELA        & Math       & ELA \\
	\hline
	$\delta$	& \multicolumn{1}{l}{FFT lower middle} & 0.146**    & -0.037     & 0.029      & 0.097      & 0.114      & -0.056 \\
	&            & (0.063)    & (0.048)    & (0.307)    & (0.188)    & (0.092)    & (0.059) \\
	   & \multicolumn{1}{l}{FFT upper middle} & 0.029      & 0.013      & -0.656*    & -0.222     & -0.035     & -0.114 \\
	&            & (0.059)    & (0.055)    & (0.386)    & (0.268)    & (0.078)    & (0.072) \\
	& \multicolumn{1}{l}{FFT high} & 0.017      & -0.006     & -1.622**   & -0.044     & -0.162     & -0.132 \\
	&            & (0.069)    & (0.060)    & (0.815)    & (0.213)    & (0.122)    & (0.086) \\
	$\eta$ & \multicolumn{1}{l}{FFT lower middle} &            &            &            &            &            &  \\
	& \multicolumn{1}{l}{\hspace{0.1cm}$\times$  baseline lower middle} &            &            & 0.111      & 0.084      & 0.083      & 0.074 \\
	&            &            &            & (0.096)    & (0.075)    & (0.095)    & (0.072) \\
	& \multicolumn{1}{l}{\hspace{0.1cm}$\times$  baseline upper middle} &            &            & 0.028      & 0.023      & -0.001     & 0.007 \\
	&            &            &            & (0.104)    & (0.087)    & (0.107)    & (0.079) \\
	& \multicolumn{1}{l}{\hspace{0.1cm}$\times$ baseline high} &            &            & 0.053      & 0.028      & 0.016      & 0.002 \\
	&            &            &            & (0.125)    & (0.107)    & (0.124)    & (0.104) \\
		    & \multicolumn{1}{l}{FFT upper middle} &            &            &            &            &            &  \\
	& \multicolumn{1}{l}{\hspace{0.1cm}$\times$  baseline lower middle} &            &            & 0.101      & 0.103      & 0.096      & 0.103 \\
	&            &            &            & (0.087)    & (0.070)    & (0.078)    & (0.069) \\
	& \multicolumn{1}{l}{\hspace{0.1cm}$\times$  baseline upper middle} &            &            & 0.001      & 0.141*     & 0.015      & 0.142* \\
	&            &            &            & (0.087)    & (0.084)    & (0.083)    & (0.080) \\
	& \multicolumn{1}{l}{\hspace{0.1cm}$\times$ baseline high} &            &            & 0.110      & 0.214**    & 0.111      & 0.222** \\
	&            &            &            & (0.115)    & (0.092)    & (0.100)    & (0.093) \\
	& \multicolumn{1}{l}{FFT high} &            &            &            &            &            &  \\
	& \multicolumn{1}{l}{\hspace{0.1cm}$\times$ baseline lower middle} &            &            & 0.269**    & 0.144*     & 0.200*     & 0.118 \\
	&            &            &            & (0.120)    & (0.075)    & (0.104)    & (0.074) \\
	& \multicolumn{1}{l}{\hspace{0.1cm}$\times$ baseline upper middle} &            &            & 0.136      & 0.232**    & 0.140      & 0.181** \\
	&            &            &            & (0.111)    & (0.089)    & (0.113)    & (0.085) \\
	& \multicolumn{1}{l}{\hspace{0.1cm}$\times$ baseline high} &            &            & 0.266**    & 0.251**    & 0.253**    & 0.182* \\
	&            &            &            & (0.130)    & (0.097)    & (0.124)    & (0.099) \\
	&            &            &            &            &            &            &  \\
	$\beta$    & \multicolumn{1}{l}{Baseline lower middle} & 0.635***   & 0.780***   & 0.517**    & 0.747***   & 0.563***   & 0.721*** \\
	&            & (0.131)    & (0.098)    & (0.195)    & (0.105)    & (0.156)    & (0.107) \\
	& \multicolumn{1}{l}{Baseline upper middle} & 1.228***   & 1.273***   & 1.143***   & 1.246***   & 1.201***   & 1.203*** \\
	&            & (0.160)    & (0.116)    & (0.218)    & (0.124)    & (0.177)    & (0.127) \\
	& \multicolumn{1}{l}{Baseline high} & 1.913***   & 2.170***   & 1.750***   & 2.121***   & 1.835***   & 2.072*** \\
	&            & (0.172)    & (0.141)    & (0.246)    & (0.148)    & (0.201)    & (0.151) \\
	
	\multicolumn{1}{c}{$\lambda$} &            &         &          & $\checkmark$        & $\checkmark$         &          & \\
	\hline
	$R^2$      &            & 0.649      & 0.592      & 0.638      & 0.589      & 0.648      & 0.593 \\
	N          &            & 8,534      & 9,641      & 8,534      & 9,641      & 8,534      & 9,641 \\
	\bottomrule
	\bottomrule
\end{tabular}%
\end{small}
	\caption*{\footnotesize \textit{Note:} The dependent variables are subject-specific test score outcomes in math and ELA. The instrumental variables are based on assigned teacher FFT (Panels A--C) and assigned peer basedline test scores (Panel B). All regressions control for the $h(x,\overline{x})$ function (see Section~\ref{sec:model}) and for randomization block fixed effects. The results for the coefficients on teacher-peer match effects (Panel B) are omitted. Analytic standard errors, clustered by randomization block, are in parentheses. FFT: framework for teaching. \\ *** significant at the 1\%-level ** significant at the 5\%-level * significant at the 10\%-level.}
	\label{tab:results_4-by-4_iv}
\end{table}


\begin{table}[htbp]
	\centering
	\caption{IV regression results of the 3 $\times$ 3 model with CLASS as measure of teaching practices. Dependent variables: student test score outcomes}
	\begin{small}
	\begin{tabular}{lrcccccc}
		\toprule
		\toprule
		&            & (1)        & (2)        & (3)        & (4)       & (5)       & (6) \\
		&            & \multicolumn{2}{c}{A. Only teacher} & \multicolumn{2}{c}{B. Full model} & \multicolumn{2}{c}{C. Without} \\
		&            & \multicolumn{2}{c}{effects} &            &            & \multicolumn{2}{c}{teacher $\times$ peer} \\
		&            & \multicolumn{2}{c}{}    &            &            & \multicolumn{2}{c}{interactions} \\
		&            & Math       & ELA        & Math       & ELA        & Math       & ELA \\
		\midrule
		$\delta$   & \multicolumn{1}{l}{CLASS middle} & 0.039      & 0.071      & -0.214     & 0.115      & 0.008      & 0.038 \\
		&            & (0.061)    & (0.067)    & (0.489)    & (0.235)    & (0.103)    & (0.072) \\
		& \multicolumn{1}{l}{CLASS high} & -0.050     & 0.010      & -0.521     & 0.085      & -0.211*    & -0.099 \\
		&            & (0.079)    & (0.088)    & (0.520)    & (0.262)    & (0.117)    & (0.107) \\
		$\eta$     & \multicolumn{1}{l}{CLASS middle } &            &            &            &            &            &  \\
		& \multicolumn{1}{l}{\hspace{0.1cm}$\times$  baseline middle} &            &            & 0.075      & 0.054      & 0.074      & 0.058 \\
		&            &            &            & (0.105)    & (0.070)    & (0.098)    & (0.067) \\
		& \multicolumn{1}{l}{\hspace{0.1cm}$\times$ baseline high} &            &            & 0.008      & 0.040      & 0.019      & 0.048 \\
		&            &            &            & (0.115)    & (0.092)    & (0.109)    & (0.091) \\
		& \multicolumn{1}{l}{CLASS high} &            &            &            &            &            &  \\
		& \multicolumn{1}{l}{\hspace{0.1cm}$\times$ baseline middle} &            &            & 0.201*     & 0.177**    & 0.198**    & 0.168** \\
		&            &            &            & (0.103)    & (0.076)    & (0.100)    & (0.077) \\
		& \multicolumn{1}{l}{\hspace{0.1cm}$\times$ baseline high} &            &            & 0.253**    & 0.157*     & 0.240**    & 0.137 \\
		&            &            &            & (0.114)    & (0.085)    & (0.106)    & (0.088) \\
		$\lambda$  & \multicolumn{1}{l}{CLASS middle} &            &            &            &            &            &  \\
		& \multicolumn{1}{l}{\hspace{0.1cm}$\times$ fraction peers middle} &            &            & 0.451      & -0.165     &            &  \\
		&            &            &            & (0.934)    & (0.473)    &            &  \\
		& \multicolumn{1}{l}{\hspace{0.1cm}$\times$ fraction peers high} &            &            & 0.153      & -0.040     &            &  \\
		&            &            &            & (0.373)    & (0.306)    &            &  \\
		& \multicolumn{1}{l}{CLASS high} &            &            &            &            &            &  \\
		& \multicolumn{1}{l}{\hspace{0.1cm}$\times$ fraction peers middle} &            &            & 0.741      & -0.168     &            &  \\
		&            &            &            & (0.916)    & (0.375)    &            &  \\
		& \multicolumn{1}{l}{\hspace{0.1cm}$\times$ fraction peers high} &            &            & 0.036      & -0.324     &            &  \\
		&            &            &            & (0.442)    & (0.382)    &            &  \\
		&            &            &            &            &            &            &  \\
		$\beta$    & \multicolumn{1}{l}{Baseline middle} & 0.929***   & 0.870***   & 0.819***   & 0.798***   & 0.820***   & 0.798*** \\
		&            & (0.095)    & (0.087)    & (0.148)    & (0.100)    & (0.143)    & (0.099) \\
		& \multicolumn{1}{l}{Baseline high} & 1.686***   & 1.645***   & 1.572***   & 1.584***   & 1.562***   & 1.586*** \\
		&            & (0.118)    & (0.115)    & (0.158)    & (0.122)    & (0.152)    & (0.124) \\
		\midrule
		$R^2$      &            & 0.607      & 0.557      & 0.603      & 0.557      & 0.607      & 0.558 \\
		N          &            & 6,320      & 6,999      & 6,320      & 6,999      & 6,320      & 6,999 \\
		\bottomrule
		\bottomrule
	\end{tabular}%
	\end{small}
	\label{tab:results_3-by-3_class}%
	\caption*{\footnotesize \textit{Note:} The dependent variables are subject-specific test score outcomes in math and ELA. The instrumental variables are based on assigned teacher CLASS (Panels A--C) and assigned peer baseline test scores (Panel B). For details on the CLASS measure of teaching practices, see Appendix~\ref{sec:data_appendix}. All regressions control for the $h(x,\overline{x})$ function (see Section~\ref{sec:model}) and for randomization block fixed effects. Analytic standard errors, clustered by randomization block, are in parentheses. \\ ***significant at the 1\%-level, **significant at the 5\%-level, *significant at the 10\%-level.
}
\end{table}%



\begin{table}[htbp]
	\centering
	\caption{Average reallocation gains in math: sensitivity to restrictions on possible assignments.  Gains expressed in test score standard deviations.}
	\begin{small}
	\begin{tabular}{lccccccccc}
		\toprule
		\toprule
		& (1)        & (2)        & (3)        & (4)        &            & (5)        & (6)        & (7)        & (8) \\
		& \multicolumn{9}{c}{Panel A. Optimal versus status quo} \\
		\cmidrule{2-10}               & \multicolumn{4}{c}{A.I Within school type}        &            & \multicolumn{4}{c}{A.II Within randomization block} \\
		& all students & high       & middle     & low        &            & all students & high       & middle     & low \\
		Gain       & 0.019      & 0.031      & 0.015      & 0.015      &            & 0.005      & 0.009      & 0.004      & 0.003 \\
		SE         & (0.007)    & (0.015)    & (0.010)    & (0.012)    &            & (0.002)    & (0.004)    & (0.003)    & (0.003) \\
		N          & 8,534      & 2,332      & 3,108      & 3,094      &            & 8,534      & 2,332      & 3,108      & 3,094 \\
		&            &            &            &            &            &            &            &            &  \\
		& \multicolumn{9}{c}{Panel B. Optimal versus worst allocation} \\
		\cmidrule{2-10}               & \multicolumn{4}{c}{B.I Within school type}        &            & \multicolumn{4}{c}{B.II Within randomization block} \\
		& all students & high       & middle     & low        &            & all students & high       & middle     & low \\
		Gain       & 0.048      & 0.089      & 0.025      & 0.041      &            & 0.011      & 0.020      & 0.008      & 0.008 \\
		SE         & (0.015)    & (0.039)    & (0.019)    & (0.028)    &            & (0.003)    & (0.009)    & (0.006)    & (0.007) \\
		N          & 8,534      & 2,332      & 3,108      & 3,094      &            & 8,534      & 2,332      & 3,108      & 3,094 \\
		\bottomrule
		\bottomrule
	\end{tabular}%
	\end{small}
	\label{tab:restrict-assignments_math}%
	\caption*{\footnotesize \textit{Note:} The table shows the average reallocation gains from implementing the optimal assignment instead of the random assignment (status quo) in Panel A, and the average reallocation gains from implementing the optimal assignment instead of the worst assignment in Panel B. The gains are expressed in test score standard deviations. The computations are based on the 3 $\times$ 3 model without teacher-by-peer interactions (see Table~\ref{tab:results_3-by-3_iv}, column 5). Panels A.I and B.I display the reallocation effects when the reallocation is carried out within school types (elementary or middle school) and across school districts. Panels A.II and B.II display the reallocation effects when the reallocation is carried out within randomization blocks. Standard errors are in parentheses and computed using the Bayesian bootstrap with 1,000 replications (see Section~\ref{sec:inference}). High/middle/low: students in the top/middle/bottom tercile of the baseline test score distribution.
	}
\end{table}%

\begin{table}[htbp]
	\centering
	\caption{Average reallocation gains in ELA: sensitivity to restrictions on possible assignments. Gains expressed in test score standard deviations}
	\begin{small}
	\begin{tabular}{lccccccccc}
		\toprule
		\toprule
		& (1)        & (2)        & (3)        & (4)        &            & (5)        & (6)        & (7)        & (8) \\
		& \multicolumn{9}{c}{Panel A. Optimal versus status quo} \\
		\cmidrule{2-10}               & \multicolumn{4}{c}{A.I Within school type}        &            & \multicolumn{4}{c}{A.II Within randomization block} \\
		& all students & high       & middle     & low        &            & all students & high       & middle     & low \\
		Gain       & 0.009      & 0.027      & 0.002      & 0.004      &            & 0.002      & 0.007      & 0.001      & 0.001 \\
		SE         & (0.007)    & (0.013)    & (0.007)    & (0.012)    &            & (0.002)    & (0.004)    & (0.002)    & (0.003) \\
		N          & 9,641      & 2,480      & 3,402      & 3,759      &            & 9,641      & 2,480      & 3,402      & 3,759 \\
		&            &            &            &            &            &            &            &            &  \\
		& \multicolumn{9}{c}{Panel B. Optimal versus worst allocation} \\
		\cmidrule{2-10}               & \multicolumn{4}{c}{B.I Within school type}        &            & \multicolumn{4}{c}{B.II Within randomization block} \\
		& all students & high       & middle     & low        &            & all students & high       & middle     & low \\
		Gain       & 0.021      & 0.066      & 0.004      & 0.006      &            & 0.005      & 0.017      & 0.001      & 0.002 \\
		SE         & (0.013)    & (0.034)    & (0.013)    & (0.022)    &            & (0.004)    & (0.009)    & (0.004)    & (0.007) \\
		N          & 9,641      & 2,480      & 3,402      & 3,759      &            & 9,641      & 2,480      & 3,402      & 3,759 \\
		\bottomrule
		\bottomrule
	\end{tabular}%
	\end{small}
	\label{tab:restrict-assignments_ela}%
		\caption*{\footnotesize \textit{Note:} The table shows the average reallocation gains from implementing the optimal assignment instead of the random assignment (status quo) in Panel A, and the average reallocation gains from implementing the optimal assignment instead of the worst assignment in Panel B. The gains are expressed in test score standard deviations. The computations are based on the 3 $\times$ 3 model without teacher-by-peer interactions (see Table~\ref{tab:results_3-by-3_iv}, column 6). Panels A.I and B.I display the reallocation gains when the reallocation is carried out within school types (elementary or middle school) and across school districts. Panels A.II and B.II display the reallocation effects when the reallocation is carried out within randomization blocks. Standard errors are in parentheses and computed using the Bayesian bootstrap with 1,000 replications. High/middle/low: students in the top/middle/bottom tercile of the baseline test score distribution.
		}
\end{table}%

\begin{landscape}
\begin{table}[htbp]
	\centering
\caption{Average reallocation gains in math: sensitivity to inclusion of teacher-by-peer interactions. Gains expressed in test score standard deviations}
\begin{small}
	\begin{tabular}{lcccccccccrcccc}
		\toprule
		\toprule
		& (1)        & (2)        & (3)        & (4)        &            & (5)        & (6)        & (7)        & (8)        &            & (9)        & (10)       & (11)       & (12) \\
		& \multicolumn{14}{c}{Panel A. Optimal versus status quo} \\
		\cmidrule{2-15}               & \multicolumn{4}{c}{A.I Within school type}        &            & \multicolumn{4}{c}{A.II Within school type and district} &            & \multicolumn{4}{c}{A.III Within randomization block} \\
		& all students & high       & middle     & low        &            & all students & high       & middle     & low        &            & all students & high       & middle     & low \\
		Gain       & 0.035      & 0.059      & 0.026      & 0.027      &            & 0.034      & 0.057      & 0.025      & 0.025      &            & 0.009      & 0.016      & 0.008      & 0.006 \\
		SE         & (0.020)    & (0.028)    & (0.017)    & (0.026)    &            & (0.022)    & (0.030)    & (0.018)    & (0.031)    &            & (0.006)    & (0.008)    & (0.006)    & (0.006) \\
		N          & 8,534      & 2,332      & 3,108      & 3,094      &            & 8,534      & 2,332      & 3,108      & 3,094      &            & 8,534      & 2,332      & 3,108      & 3,094 \\
		&            &            &            &            &            &            &            &            &            &            &            &            &            &  \\
		& \multicolumn{14}{c}{Panel B. Optimal versus worst allocation} \\
		\cmidrule{2-15}               & \multicolumn{4}{c}{B.I Within school type}        &            & \multicolumn{4}{c}{B.II Within school type and district} &            & \multicolumn{4}{c}{B.III Within randomization block} \\
		& all students & high       & middle     & low        &            & all students & high       & middle     & low        &            & all students & high       & middle     & low \\
		Gain       & 0.098      & 0.170      & 0.070      & 0.073      &            & 0.085      & 0.146      & 0.061      & 0.064      &            & 0.022      & 0.035      & 0.019      & 0.015 \\
		SE         & (0.048)    & (0.093)    & (0.043)    & (0.053)    &            & (0.047)    & (0.089)    & (0.046)    & (0.054)    &            & (0.012)    & (0.018)    & (0.013)    & (0.012) \\
		N          & 8,534      & 2,332      & 3,108      & 3,094      &            & 8,534      & 2,332      & 3,108      & 3,094      &            & 8,534      & 2,332      & 3,108      & 3,094 \\
		\bottomrule
		\bottomrule
	\end{tabular}%
	\end{small}
	\label{tab:teach-by-peer_math}%
		\caption*{\footnotesize \textit{Note:} The table shows the average reallocation gains from implementing the optimal assignment  instead of the random assignment (status quo) in Panel A, and the average reallocation gains from implementing the optimal assignment instead of the worst assignment in Panel B. The gains are expressed in test score standard deviations. The computations are based on the 3 $\times$ 3 model with teacher-by-peer interactions (see Table~\ref{tab:results_3-by-3_iv}, column 3). Columns 1--4 display the reallocation effects when the reallocation is carried out within school types (elementary or middle school) and across districts, columns 5--8 display the reallocation effects when the reallocation is carried out within school types and districts, and columns 9--12 display the reallocation effects when the reallocation is carried out within randomization blocks. Standard errors are in parentheses and computed using the Bayesian bootstrap with 1,000 replications. High/middle/low: students in the top/middle/bottom tercile of the baseline test score distribution.}
\end{table}%
\end{landscape}

\begin{landscape}
\begin{table}[htbp]
	\centering
	\caption{Average reallocation gains in ELA: sensitivity to inclusion of teacher-by-peer interactions. Gains expressed in test score standard deviations}
	\begin{small}
	\begin{tabular}{lcccccccccrcccc}
		\toprule
		\toprule
		& (1)        & (2)        & (3)        & (4)        &            & (5)        & (6)        & (7)        & (8)        &            & (9)        & (10)       & (11)       & (12) \\
		& \multicolumn{14}{c}{Panel A. Optimal versus status quo} \\
		\cmidrule{2-15}               & \multicolumn{4}{c}{A.I Within school type}        &            & \multicolumn{4}{c}{A.II Within school type and district} &            & \multicolumn{4}{c}{A.III Within randomization block} \\
		& all students & high       & middle     & low        &            & all students & high       & middle     & low        &            & all students & high       & middle     & low \\
		Gain       & 0.044      & 0.042      & 0.043      & 0.045      &            & 0.039      & 0.034      & 0.040      & 0.042      &            & 0.012      & 0.011      & 0.011      & 0.014 \\
		SE         & (0.027)    & (0.023)    & (0.024)    & (0.044)    &            & (0.023)    & (0.019)    & (0.020)    & (0.037)    &            & (0.008)    & (0.005)    & (0.007)    & (0.011) \\
		N          & 9,641      & 2,480      & 3,402      & 3,759      &            & 9,641      & 2,480      & 3,402      & 3,759      &            & 9,641      & 2,480      & 3,402      & 3,759 \\
		&            &            &            &            &            &            &            &            &            &            &            &            &            &  \\
		& \multicolumn{14}{c}{Panel B. Optimal versus worst allocation} \\
		\cmidrule{2-15}               & \multicolumn{4}{c}{B.I Within school type}        &            & \multicolumn{4}{c}{B.II Within school type and district} &            & \multicolumn{4}{c}{B.III Within randomization block} \\
		& all students & high       & middle     & low        &            & all students & high       & middle     & low        &            & all students & high       & middle     & low \\
		Gain       & 0.091      & 0.070      & 0.080      & 0.116      &            & 0.080      & 0.061      & 0.070      & 0.101      &            & 0.027      & 0.023      & 0.021      & 0.035 \\
		SE         & (0.052)    & (0.037)    & (0.043)    & (0.085)    &            & (0.044)    & (0.031)    & (0.035)    & (0.069)    &            & (0.016)    & (0.011)    & (0.012)    & (0.026) \\
		N          & 9,641      & 2,480      & 3,402      & 3,759      &            & 9,641      & 2,480      & 3,402      & 3,759      &            & 9,641      & 2,480      & 3,402      & 3,759 \\
		\bottomrule
		\bottomrule
	\end{tabular}%
	\end{small}
	\label{tab:teach-by-peer_ela}%
		\caption*{\footnotesize \textit{Note:} The table shows the average reallocation gains from implementing the optimal assignment  instead of the random assignment (status quo) in Panel A, and the average reallocation gains from implementing the optimal assignment instead of the worst assignment in Panel B. The gains are expressed in test score standard deviations. The computations are based on the 3 $\times$ 3 model with teacher-by-peer interactions (see~Table~\ref{tab:results_3-by-3_iv}, column 4). Columns 1--4 display the reallocation effects when the reallocation is carried out within school types (elementary or middle school) and across districts, columns 5--8 display the reallocation effects when the reallocation is carried out within school types and districts, and columns 9--12 display the reallocation effects when the reallocation is carried out within randomization blocks. Standard errors are in parentheses and computed using the Bayesian bootstrap with 1,000 replications. High/middle/low: students in the top/middle/bottom tercile of the baseline test score distribution.}
\end{table}%
\end{landscape}

\clearpage

\section{Data Appendix}
\label{sec:data_appendix}

\setcounter{figure}{0} \renewcommand{\thefigure}{B.\arabic{figure}} 
\setcounter{table}{0} \renewcommand{\thetable}{B.\arabic{table}}

\subsection{Construction of the dataset from the MET files}

Our dataset combines eight different data files from the MET study (2018 release): the randomization file, the teacher file, the class section file, the student file, two classroom observation score files (the CLASS and the FFT file), as well as the district-wide files for the school years 2009/10 and 2010/11.  

The basis of our data construction is the randomization file. It contains identifiers for all students who were randomly assigned to a teacher in the second year of the MET study, and an identifier (identifiers) for their assigned teacher(s). Through the teacher identifiers, we merge this dataset to the FFT and CLASS files, and thus obtain the FFT and CLASS of the assigned teachers. Moreover, we use the teacher identifiers to merge the data to the teacher file, in order to obtain teacher background characteristics of the assigned teachers.

Student characteristics and test score outcomes come from the student file and the district-wide files. Through a student identifier, we first merge the randomization file with the student file, which contains individual information for all students who were part of the MET study and still had a MET teacher at the end of the school year; i.e., the student file does not contain any information on students who switch to a non-MET teacher within the same school, a non-MET school, or a non-MET district. We use the student file to extract students' demographic characteristics, baseline test scores (test scores in school year 2009/10), and test score outcomes (test scores in school year 2010/11) in math and ELA. While there are no missings for student background characteristics, some of the students have missing baseline test scores or missing test score outcomes. Therefore, we obtain the missing test score information from the district-wide files. 

To construct information on the peer group composition in the assigned classroom, we average the student background characteristics and baseline test scores at the level of the assigned teacher and randomization block, since each teacher can only be assigned to one classroom within a randomization block. To be precise, we construct the leave-own-out mean, i.e., the classroom mean excluding the student herself.

Information on the realized teacher and the realized peers are constructed based on the student file. This file contains an identifier for the section that the student attended in school year 2010/11, as well as an identifier for the teacher who taught the section. Based on this teacher identifer, we merge information on the FFT, CLASS, and background characteristics of the realized teacher. Furthermore, we construct leave-own-out means of peer characteristics based on the section identifiers. We add information on the size of the realized classroom from the class section file.

\subsection{Construction of the estimation sample}

In our analysis, we focus on students in grades 4-8 who were randomized to a teacher before the start of the school year. We create separate samples for math and ELA. In total, the randomization sample contains information on about 16,000 students who were randomized to a teacher in math and ELA in grades 4-8.

\begin{table}[H]
	\caption{Sample construction from the randomization file}
\begin{small}
\begin{tabular}{cp{19.57em}rrrrr}
	\toprule
	\toprule
	& \multicolumn{1}{r}{} & \multicolumn{1}{c}{(1)} & \multicolumn{1}{c}{(2)} &            & \multicolumn{1}{c}{(3)} & \multicolumn{1}{c}{(4)} \\
	\midrule
	\multicolumn{2}{l}{Restrictions} & \multicolumn{2}{c}{Math} &            & \multicolumn{2}{c}{ELA} \\
	\cmidrule{3-4}\cmidrule{6-7}           & \multicolumn{1}{r}{} & \multicolumn{1}{c}{N} & \multicolumn{1}{p{3.57em}}{percent} &            & \multicolumn{1}{c}{N} & \multicolumn{1}{p{3.57em}}{percent} \\
	& \multicolumn{1}{r}{} &            & \multicolumn{1}{p{3.57em}}{of base} &            &            & \multicolumn{1}{p{3.57em}}{of base} \\
	& \multicolumn{1}{r}{} &            & \multicolumn{1}{p{3.57em}}{sample} &            &            & \multicolumn{1}{p{3.57em}}{sample} \\
	1.         & Students in randomization file & 15,749     & -          &            & 16,252     & - \\
	2.         & Record in the student file & 10,268     & -          &            & 11,271     & - \\
	3.         & At least two classrooms per randomization block (base sample) & 9,824      & 100\%      &            & 10,856     & 100\% \\
	4.         & Baseline test scores and test score outcomes available & 9,245      & 94\%       &            & 10,136     & 93\% \\
	5.         & FFT of the assigned teacher available & 9,066      & 92\%       &            & 10,057     & 93\% \\
	6.         & FFT of the realized teacher available & 8,724      & 89\%       &            & 9,767      & 90\% \\
	7.         & Information on the assigned peers available & 8,718      & 89\%       &            & 9,762      & 90\% \\
	8.         & Information on the realized peers available  & 8,717      & 89\%       &            & 9,761      & 90\% \\
	9.         & At least two classrooms per randomization block after applying all restrictions & 8,534      & 87\%       &            & 9,641      & 89\% \\
	\bottomrule
	\bottomrule
\end{tabular}%
\end{small}
	\caption*{\footnotesize \textit{Note:} Sample restrictions used to create the estimation sample from MET data files, school year 2010/11.}
	\label{tab:restrictions}
\end{table}

Table~\ref{tab:restrictions} details further restrictions that we apply to construct our estimation dataset. We require each student to have a record in the student file, since we use this file to identify the realized teacher, the realized peer group, and the student background variables. About two-thirds of the students in the randomization sample can be identified in the student file. Moreover, we restrict our sample to only randomization blocks with at least two classrooms. The resulting dataset forms our base sample. 

After constructing our base sample, we remove observations with missing information on baseline test scores or test score outcomes, with missing information on the FFT of the assigned or realized teacher, and with missing information on the baseline test scores of the assigned or realized peers. We further remove all randomization blocks with only one classroom after applying these restrictions. Our resulting estimation sample contains information on about 8,500 students in math and 9,600 students in ELA. Thus, we retain about 87 percent of observations from the base sample in math, and 89 percent of observations from the base sample in ELA.

Our sample size is close to the sample size reported by~\citet{GarrettSteinberg2015}. We obtain a slightly larger sample size because we complete missing test score information in the student file with information from the district-wide files. These files are part of the 2018 MET release and were not available when \citet{GarrettSteinberg2015} published their study.

Our estimation dataset does not contain any missing information for the main variables used in the analysis; however, it contains some missings in teacher and student demographics. With the exception of free/reduced-price lunch eligibility (30 percent missings in math and 25 percent missings in ELA), the student background variables contain (virtually) no missings. Further missings occur in the teacher demographics: teacher gender and race/ethnicity are missing for 3--4 percent of the estimation sample. Teacher experience and teachers' education are missing for about 30 percent of the estimation sample each, because one school district did not provide information on teachers' education, and another district did not provide information on teachers' experience. Full information on teacher demographics is only available for 42 percent of the math sample, and for 48 percent of the ELA sample.

\subsection{Sample comparisons}

This section investigates in how far the randomization sample differs from the estimation sample. It compares the distribution of student baseline test scores and teacher FFT in the original randomization sample with the distribution in the estimation sample. 

\begin{figure}[h!tbp]
	\centering
	\caption{Distribution of baseline test scores in the randomization sample and the estimation sample}
	\includegraphics[width=.8\linewidth]{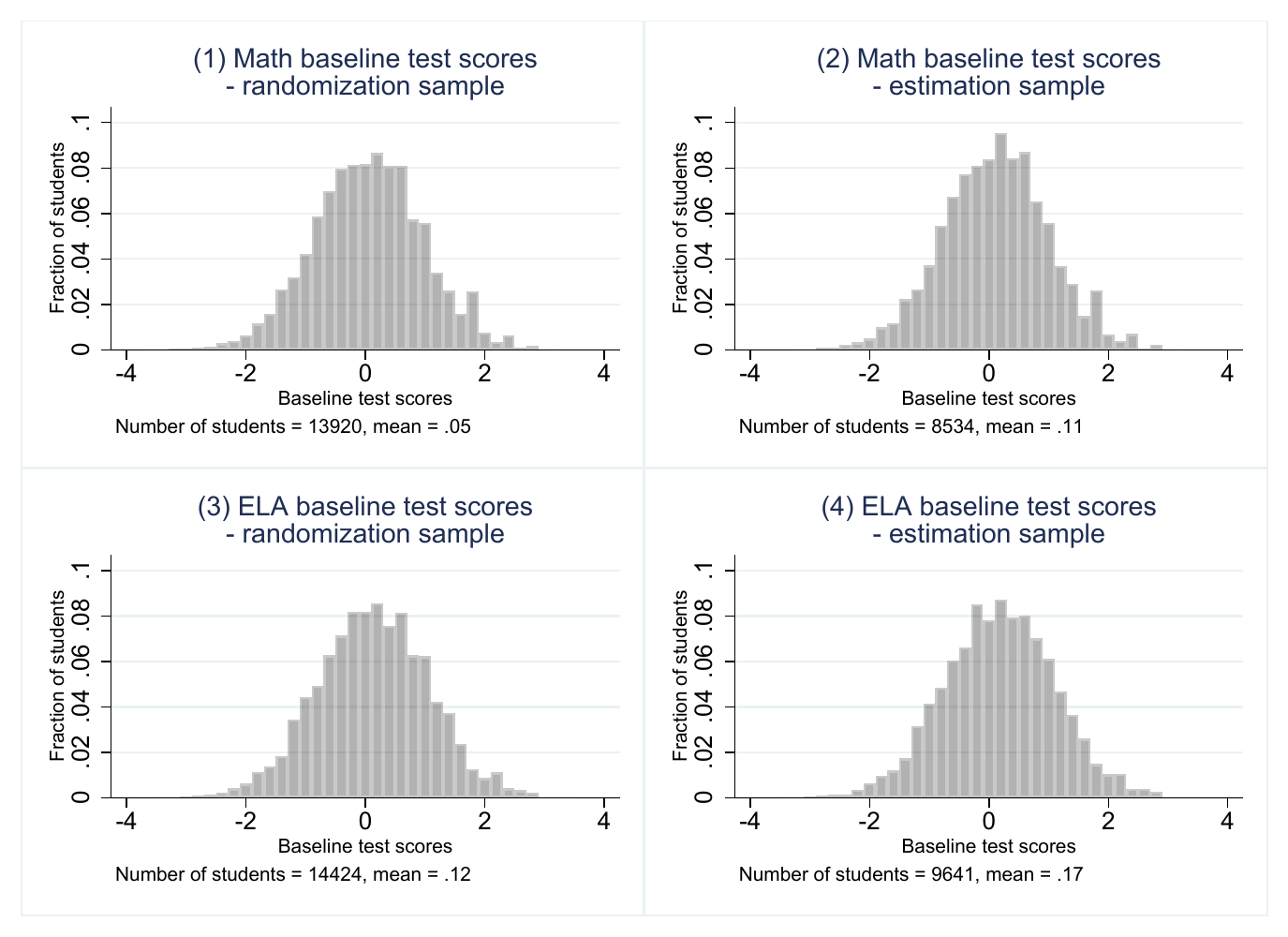}
	\label{fig:compare-baseline-testscores}
	\caption*{\footnotesize \textit{Note:} The figure compares the distributions of student baseline test scores across the randomization sample and the estimation sample for math (Panels 1 and 2) and ELA (Panels 3 and 4).}
\end{figure}

The randomization sample consists of about 15,700 students in math and 16,300 students in ELA; baseline test scores are available for about 13,900 students in math and for about 14,400 students in ELA (see Figure~\ref{fig:compare-baseline-testscores}). The remaining student observations can neither be matched to the district-wide files nor be matched to the student file. Our estimation sample contains only students in the student file \textendash{} because we use this file to identify realized teachers and peers \textendash{} and is thus considerably smaller (about 8,500 students in math and about 9,600 students in ELA). Students in the estimation dataset have on average higher baseline test scores than the students in the randomization dataset. The difference amounts to 0.06 standard deviations in math and to 0.05 standard deviations in ELA. 

\begin{figure}[h!tbp]
	\centering
	\caption{Distribution of FFT in the randomization sample and the estimation sample}
	\includegraphics[width=.8\linewidth]{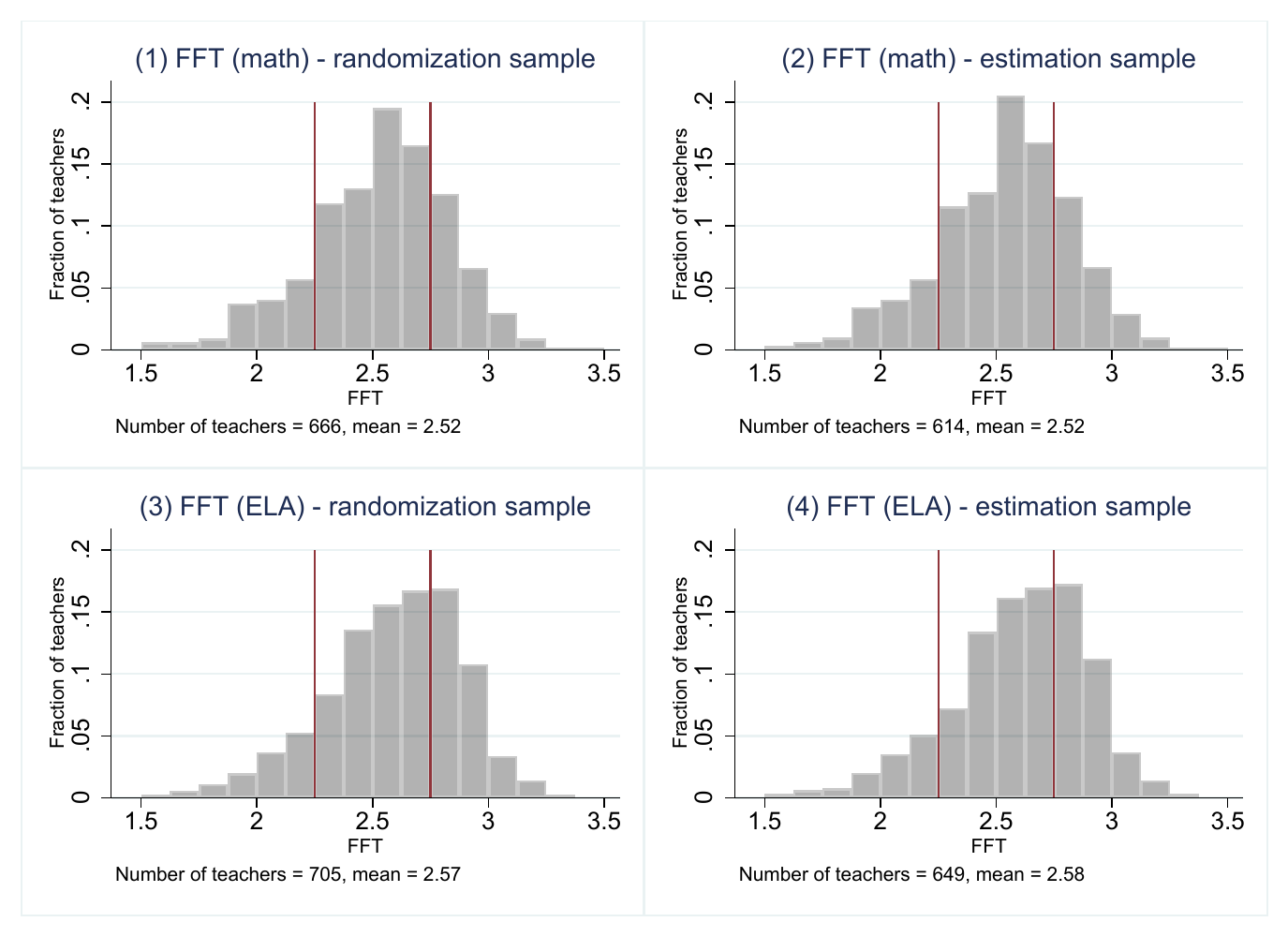}
	\label{fig:compare-fft}
	\caption*{\footnotesize \textit{Note:} The figure compares the distributions of teacher FFT across the randomization sample and the estimation sample for math teachers (Panels 1 and 2) and ELA teachers (Panels 3 and 4). The red lines denote the FFT cutoffs we use to classify teachers in our preferred specification.}
\end{figure}

Teacher FFT does not differ appreciably between the randomization sample and the estimation sample (see Figure~\ref{fig:compare-fft}). In the math randomization sample for grades 4--8, 666 teachers have non-missing information on FFT; the estimation sample includes 614 teachers. In both samples, the average FFT is 2.52. In the ELA randomization sample for these grades, 705 teacher have non-missing FFT, and the estimation sample contains 649 teachers. The average FFT in both samples are nearly identical with 2.57 and 2.58. 

\subsection{Information on the CLASS measure}
\label{sec:class}

\paragraph{Information from the MET User Guide.} The following information about the CLASS (Classroom Assessment Scoring System) protocol is taken from the MET User Guide (\citealp{WhiteRowan2018}, p. 32--33):

``CLASS is an observational protocol designed to measure the extent to which teachers effectively support children's social and academic development. Two different versions of CLASS were used in the MET Study: the Upper Elementary (Grades 4-5) and the Secondary (Grades 6-9).

``The CLASS instrument is divided into three broad domains of measurement: Emotional Support,
Classroom Organization, and Instructional Support. Each domain, in turn, is measured by a number
of dimensions. The domain ``Emotional Support,'' for example, refers to the emotional tone in a
classroom, which can be measured along four dimensions: positive climate, negative climate, teacher sensitivity, and regard for student perspectives. The domain ``Classroom Organization'' refers to the ways a classroom is structured to manage students' behavior, time, and attention, which can be measured along three dimensions: behavior management, productivity, and instructional learning formats. The domain ``Instructional Supports'' refers to the ways a teacher provides supports to encourage student conceptual understanding and student problem solving and can be measured along four dimensions: content understanding, analysis and problem solving, instructional dialogue, and quality of feedback. [...]

``CLASS scoring is done using a detailed scoring rubric. In this rubric, a classroom is scored on each instructional dimension at 15-minute intervals using a 7-point scale. For the MET Study, only the first 30 minutes of each video was scored. Scores are assigned based on anchor descriptions of what is to be observed in order for a classroom to be scored at ``high,'' ``mid,'' and ``low'' points on the 7-point scale. In the MET Study, dimension scores are often aggregated to higher levels of analysis simply by averaging raters' scores to get a single segment score and then calculating the harmonic mean of segment scores across all segments for a particular target of measurement (e.g., a day, a class section, a teacher). Standard errors of measurement for these derived scores are not generally reported.''

\paragraph{Use of the CLASS measure in our study.} We use the CLASS measure to test the sensitivity of our results to the classroom observation protocol used. To construct a unique measure for each teacher, we take the average across the three CLASS domains. 

Our estimation sample contains 466 teachers in the math sample and 495 teachers in the ELA sample. In the math sample, CLASS ranges from 2.54 to 5.58, with a mean of 4.34. In the ELA sample, CLASS ranges from 2.95 to 5.58, with a mean of 4.39.  We split teachers in three categories according to this measure. We choose the cutoff values of 4 and 4.5 to carry out the split. Both in the math and in the ELA sample, 20 percent of the teachers are classified as low, 42 percent as middle, and 38 percent as high.

\clearpage

\section{Optimal allocation: linear program}
\label{sec:optimal-compute}

\setcounter{figure}{0} \renewcommand{\thefigure}{C.\arabic{figure}} 
\setcounter{table}{0} \renewcommand{\thetable}{C.\arabic{table}}

The linear program aims at maximizing the aggregate test score outcomes in the data. We use $\widehat{{Y}}_{c}(w)$ to denote the predicted aggregate outcome of classroom $c$ when assigned a teacher of level $w$, where $w = w_L$ when assigned a low-, $w = w_M$ when assigned a middle-, and $w = w_H$ when assigned a high-FFT teacher (see Section~\ref{sec:optimal} for details). We use $C$ to denote the total number of teachers in the dataset, and $C_w$ to denote the number of teachers of level $w$ in the dataset. We define an indicator variable $\alpha_{cw}$, which takes the value 1 if classroom $c$ is taught by a teacher of level $w$, and 0 otherwise. We also define an assignment matrix $A$, which contains all $\alpha_{cw}$. The linear program can then be written as:

\begin{equation*}
\underset{A}{\text{max}} \sum_{c = 1}^{C} \sum_{w \in \{w_L,w_M,w_H\}} \alpha_{cw} \widehat{Y}_{c}(w)
\end{equation*}

subject to

\begin{eqnarray*}
	\sum_{w \in \{w_L,w_M,w_H\}} \alpha_{cw} = 1 \quad \forall \quad c \in C
\end{eqnarray*}

\begin{eqnarray*}
	\sum_{c = 1}^C \alpha_{cw} = C_w \quad \text{for} \ w \in \{w_L,w_M,w_H\}
\end{eqnarray*}

\begin{eqnarray*}
	\alpha_{cw} \in \{0, 1\}
\end{eqnarray*}

This is a transportation problem with $C + 3 + (3 \times C)$ constraints. We solve the transportation problem in \textit{R} using \tt{lpSolve}.

\end{document}